\documentclass{emulateapj}
\usepackage{amsmath}

\begin{document}

\title{Giant H$\alpha$ nebula surrounding the starburst merger NGC~6240\altaffilmark{1}}

\author{
Michitoshi Yoshida\altaffilmark{2},
Masafumi Yagi\altaffilmark{3,7},
Youichi Ohyama\altaffilmark{4},
Yutaka Komiyama\altaffilmark{3,8},
Nobunari Kashikawa\altaffilmark{3,8},
Hisashi Tanaka\altaffilmark{5},
and
Sadanori Okamura\altaffilmark{6,7},
}
\email{yoshidam@hiroshima-u.ac.jp}


\altaffiltext{1}{Based on data collected at the Subaru Telescope, which is operated by the
National Astronomical Observatory of Japan.}

\altaffiltext{2}{Hiroshima Astrophysical Science Center, Hiroshima University,
Higashi-Hiroshima, Hiroshima 739-8526, Japan}

\altaffiltext{3}{Optical and Infrared Astronomy Division, National 
Astronomical Observatory,
Mitaka, Tokyo 181-8588, Japan}

\altaffiltext{4}{Institute of Astronomy and Astrophysics, Academia Sinica, 
11F of Astronomy-Mathematics Building, AS/NTU
No.1, Sec. 4, Roosevelt Rd, Taipei 10617, Taiwan, R.O.C.}

\altaffiltext{5}{Department of Physical Science, Hiroshima University,
Higashi-Hiroshima, Hiroshima 739-8526, Japan}

\altaffiltext{6}{Department of Advanced Sciences, Faculty of Science and Engineering,
Hosei University, Koganei, Tokyo 184-8584, Japan}

\altaffiltext{7}{Graduate School of Science and Engineering, Hosei University,
Koganei, Tokyo, 184-8584, Japan}

\altaffiltext{8}{The Graduate University for Advanced Studies (SOKENDAI), 
Mitaka, Tokyo 181-8588, Japan}


\objectname[NGC 6240]{}


\begin{abstract}

We revealed the detailed structure of a vastly extended H$\alpha$-emitting nebula
(``H$\alpha$ nebula'') 
surrounding the starburst/merging galaxy NGC~6240 by deep
narrow-band imaging observations
with the Subaru Suprime-Cam.
The extent of the nebula is $\sim$90 kpc in diameter and the total H$\alpha$ luminosity
amounts to $L_{\rm H\alpha} \approx 1.6 \times 10^{42}$ erg~s$^{-1}$.
The volume filling factor and the mass of the warm ionized gas are 
$\sim$10$^{-4}$--10$^{-5}$ and $\sim$$5 \times 10^8$ $M_\odot$, respectively.
The nebula has a complicated structure, 
which includes numerous filaments, loops, bubbles, and knots.
We found that there is a tight spatial correlation between the H$\alpha$ nebula and the 
extended soft X-ray-emitting gas, both in large and small scales.
The overall morphology of the nebula is dominated by filamentary structures radially extending
from the center of the galaxy.
A large-scale bi-polar bubble extends along the minor axis of the main stellar disk.
The morphology strongly suggests that the nebula was formed 
by intense outflows -- superwinds -- 
driven by starbursts. 
We also found three bright knots embedded in a looped filament of ionized gas that 
show head-tail morphologies 
in both emission-line and continuum, suggesting close interactions between the outflows 
and star forming regions.
Based on the morphology and surface brightness distribution of the H$\alpha$ nebula,
we propose the scenario
that three major episodes of starburst/superwind activities which
were initiated $\sim$10$^2$ Myr ago formed the extended ionized gas nebula of NGC~6240.

\end{abstract}




\section{Introduction}

Galaxy mergers lead to intense star formation, i.e., starburst
\citep{Mihos1994,Barnes1998,Schweizer1998,Genzel2001,Hopkins2008}.
The gravitational perturbations and dynamical disturbances induced by the merging process
remove angular momentum from the galactic interstellar medium (ISM), 
The ISM thus settles into the gravitational center of the system
leading to starburst \citep{Teyssier2010,Hopkins2013}.
The collective effects of supernovae and stellar winds from massive stars within the 
starburst region drive large-scale outflows, or ``superwinds'' \citep[e.g.,][]{Veilleux2005}. 
Galaxy mergers and their successive processes, starbursts and superwinds, 
significantly alter the morphology, 
stellar population, gas content, and chemical abundance of the galaxy.
Together with statistical investigations based on a large sample of mergers, 
detailed case studies of merging starburst galaxies provide us with 
important clues for understanding the above process and its impact on galaxy evolution. 

NGC~6240 is a well studied merger/starburst galaxy in the local universe
\citep[$z \approx 0.0245$;][]{Downs1993}.
It has a highly disturbed morphology indicating that two spiral galaxies are 
in the process of colliding with each other.
NGC~6240 is often considered as an ultra-luminous infrared galaxy (ULIRG) in the 
local universe, although its infrared luminosity 
\citep[$\approx$$7\times10^{11}$ $L_{\odot}$:][]{Sanders1996} is 
slightly smaller than that of the criterion for ULIRG 
\citep[$L_{\rm IR}$(8 -- 1000 $\mu$m)$>10^{12}$ $L_{\odot}$;][]{Lonsdale2006}.
This system includes double nuclei (the northern and southern nuclei; hereafter referred to as 
the n-nucleus and the s-nucleus)  
whose separation is $\sim$1 kpc.
These nuclei are hard X-ray point sources suggesting that these are the colliding 
active galactic nuclei (AGNs) of 
the progenitor galaxies \citep{Komossa2003}. 
The s-nucleus is $\sim$3 times brighter than the n-nucleus in the 0.2--10 keV band
\citep{Komossa2003}.

It is well known that a bright optical emission line nebula is associated with NGC~6240.
Early studies by various authors \citep[e.g.][]{Heckman1987,Armus1990,Keel1990}
suggested that this optical nebula is primarily excited by 
shock-heating induced by a starburst superwind.
High spatial resolution {\it Chandra} observations have revealed that the 
central H$\alpha$ nebula (``butterfly nebula'') is spatially
coincident with the soft X-ray-emitting hot gas surrounding the double nuclei
\citep{Komossa2003,Gerssen2004}.
 
In addition, there is a highly extended ionized gas region surrounding NGC~6240.
\citet{Heckman1987,Heckman1990} detected faint H$\alpha$+[\ion{N}{2}] emission filaments
extending $\gtrsim$30 kpc from the center.
\citet{Veilleux2003} revealed a more extended optical emission line nebula
approximately 80 kpc in diameter.
An extended hot gas nebula of NGC~6240 has also been found in soft X-rays
\citep{Komossa2003,Huo2004}.
A spatial correlation between the optical emission line nebula and
the extended soft X-ray gas 
was identified by \citet{Veilleux2003}. 
Recently, \citet{Nardini2013} showed the extent and detailed structure of the very
extended (its total size reaches $\sim$100 kpc) soft X-ray-emitting gas.

Although the central region of NGC~6240 has been studied many times, the detailed
structure of the extended optical emission line gas is not yet understood.
Here we present the results of deep optical narrow-band imaging of NGC~6240
using the Subaru Suprime-Cam.
Our deep observations are the first to reveal the detailed structure of the extended H$\alpha$
emitting warm ionized gas surrounding this galaxy. 

We assumed cosmological parameters of
($h_0$, $\Omega_m$, $\Omega_\lambda$) = (0.705, 0.27, 0.73)  \citep{Komatsu2011}
and that the luminosity distance of NGC~6240 is 107 Mpc.
The linear scale at the galaxy is 492 pc arcsec$^{-1}$ under this 
assumption.

\section{Observations}

\begin{deluxetable}{lccc}
  \tablewidth{0pt}
  \tablecaption{Journal of the observations}
  \tablehead{
    \colhead{Filter ID} & 
    \colhead{PSF size} &
    \colhead{SB$_{\rm lim}$ \tablenotemark{a}} &
    \colhead{Exposure} \\
    \colhead{} &
    \colhead{[arcsec]} &
    \colhead{[ABmag~arcsec$^{-2}$]} &
    \colhead{}
  }
  \startdata
N-A-L671 & 0.88 & 27.5 & 10 $\times$ 600 s  \\
$B$         & 0.91 & 28.5 &  5 $\times$ 300 s \\
$R_{\rm C}$ & 0.76 & 27.7 & 5 $\times$ 180 s  \\
$i^\prime$ & 0.87 & 27.3 &  5 $\times$ 180 s\\ 
\enddata
  \tablenotetext{a}{1$\sigma$ fluctuation of sky flux in an aperture of 2 arcsec diameter.}

\label{tab:obs}
\end{deluxetable}

Narrow-band H$\alpha$+[\ion{N}{2}] and broad-band $B$, $R_{\rm C}$, and $i^\prime$ imaging 
observations of NGC~6240 were made using the Suprime-Cam \citep{Miyazaki2002} attached to 
the Subaru Telescope on April 30 (UT) 2014.
A journal of the observations is shown in Table \ref{tab:obs}.
We used the N-A-L671 narrow-band filter \citep{Yagi2007,Yoshida2008}
for H$\alpha$+[\ion{N}{2}] imaging of the galaxy ($z \approx 0.0245$).
The central wavelength and full-width at half maximum (FWHM) of the
transmission curve of the filter were 6712 \AA\ and 120 \AA\, respectively.
The total exposure time was 6,000 sec for N-A-L671.
The sky was clear and photometric.
Seeing was 0\farcs 7--0\farcs 9 on the observing night (Table \ref{tab:obs}).

\section{Data Reduction}

Data reduction was carried out in a standard manner.
Overscan subtraction, flat-fielding, distortion correction, sky subtraction, and
mosaicking were performed with custom-made Suprime-Cam data-reduction
software, nekosoft \citep{Yagi2002}.
During the overscan subtraction procedure, we also conducted cross-talk 
correction and gain correction
between the CCDs \citep[see][]{Yagi2012}.
We used sky flats, which had been created from a number of object frames taken
during the same
observing run, for the flat-fielding correction of the $B$ and $R_{\rm C}$ band frames.
For the N-A-L671 and $i^\prime$ frames, dome flat data were used for flat-fielding.
We then extracted an area of 400\arcsec\ $\times$ 400\arcsec\ (197 kpc $\times$ 197 kpc) 
centered on the s-nucleus of NGC~6240 from each flat-fielded frame.

\begin{figure*}
\plotone{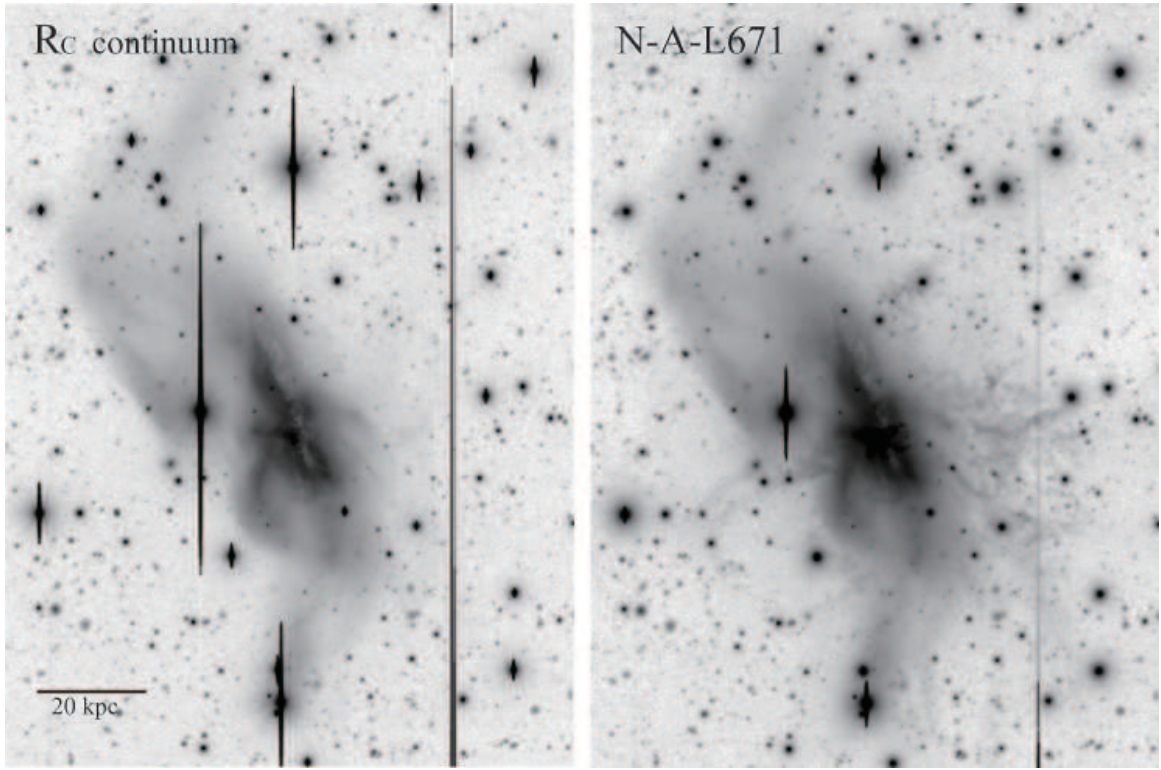}
\caption{Left panel: Emission-line-free $R_{\rm C}$ band image of NGC~6240. 
Right panel: The H$\alpha$ narrow-band (N-A-L671) image of NGC~6240. 
North is up and east is left.
Extended complex H$\alpha$-emitting filaments can be seen in the east and
west sides of the galaxy.
}
\label{fig:R-H-alpha}
\end{figure*}

After adjusting the positions and scales, the frames for each band were combined
using a 3 sigma clipping method.
The flux calibration was performed using the 9th Data Release of the 
Sloan Digital Sky Survey (SDSS) photometric
catalog \citep{Ahn2012}.
The details of the flux calibration are described in Appendix A and 
also in \citet{Yagi2013}.
The FWHM of the point-spread function (PSF) was 
measured by fitting a two-dimensional (2D) Gaussian
function to the brightness profiles of non-saturated star images for each band.
The limiting surface brightness was estimated by photometry using randomly-sampled 
2 arcsec apertures on the blank regions.
The PSF FWHMs and limiting magnitudes of the final $B$, $R_{\rm C}$, $i^\prime$
and N-A-L671 images are listed in Table \ref{tab:obs}. 
The reduced N-A-L671 image is shown in the right panel of Figure \ref{fig:R-H-alpha}.

To obtain a net (pure) H$\alpha$+[\ion{N}{2}] image of NGC~6240, we subtracted
the scaled $R_{\rm C}$ frame from the N-A-L671 frame.
Scaling was performed using several unsaturated field stars common to both frames.
Before subtraction, we convolved 2D double Gaussian
functions to the $R_{\rm C}$ frame to adjust the PSFs of both of the frames.
After subtraction, we multiplied the net H$\alpha$+[\ion{N}{2}] image by a factor of 1.23
to correct the over-subtraction of the H$\alpha$+[\ion{N}{2}] flux due to the contamination of
the emission lines in the $R_{\rm C}$ frame (see Appendix B).

In addition, we subtracted the net H$\alpha$+[\ion{N}{2}] that was multiplied by a factor of 0.23 from 
the scaled $R_{\rm C}$ frame to create the emission-line-free $R_{\rm C}$ frame. 
The emission-line-free $R_{\rm C}$ image is shown in the left panel of Figure \ref{fig:R-H-alpha}.
The net H$\alpha$+[\ion{N}{2}] image is shown in Figure \ref{fig:H-alpha} 
using three different contrasts. 

The H$\alpha$+[\ion{N}{2}] image thus created is affected by the positional variation
of the emission line ratios in the nebula.
We assumed constant emission line ratios over the entire region of the H$\alpha$ nebula in
the above procedure.
However, it is well known that there is a tendency for the flux ratios of 
the forbidden lines to the H$\alpha$ emission
line to be large at larger distances from the center of NGC~6240
\citep{Heckman1990,Keel1990,Veilleux2003}.
This causes the systematic underestimation of the net H$\alpha$+[\ion{N}{2}] flux 
at the outer region of the nebula.
In addition, the combination of the recession velocity variation of 
the nebula and the transmission curve of the
N-A-L671 filter affects the detected H$\alpha$+[\ion{N}{2}] flux.
The error of the H$\alpha$+[\ion{N}{2}] flux due to this effect was estimated 
to be  $\approx$$\pm$5\%;
the variation of the transmission of the filter is $\approx$5\%\ in
the recession velocity range of the ionized gas nebula of NGC~6240 
\citep{Heckman1990,Keel1990,Gerssen2004}.

\section{Results}

\subsection{Main stellar disk}

The morphology of NGC~6240 in optical broad bands (e.g. $R_{\rm C}$ band image: Left panel of Figure \ref{fig:R-H-alpha})
is complicated, but there are some characteristic features in it.
The bright part of the system is elongated as long as $\sim$20 kpc with a position angle
(PA) of $\sim$30$^{\circ}$ and 
the clear dust lane runs along this structure.
In the southeastern side of this bright feature, a ``banana''-shaped tail whose PA is $\sim$$-20^{\circ}$ is connected to the center.
A ``S''-shaped extended faint tidal feature surrounds these bright parts of NGC~6240 (Figure \ref{fig:R-H-alpha}).

From a morphological point of view, we suspect that the elongated bright part would be the remnant of the
disk of the primary galaxy of this colliding system.
We tried to extract the disk component from the central bright part by fitting an exponential disk
to the $i^\prime$ band image.
We used GALFIT \citep{Peng2002,Peng2010} for profile fitting.
We added four curved S$\acute{\rm e}$rsic components to fit residuals of the exponential disk fitting. 
In the fitting procedure, we excluded the circular region whose radius is 3\arcsec at the center to avoid
the influence of saturation and AGN components.
In addition, we also excluded bright stars and blooming patterns around the galaxy.

\begin{deluxetable}{cccc}
  \tablewidth{0pt}
  \tablecaption{Exponential disk parameters}
  \tablehead{
    \colhead{$i^\prime$ magnitude \tablenotemark{a}} & 
    \colhead{$R_s$ \tablenotemark{b}} &
    \colhead{PA}  &
    \colhead{inclination \tablenotemark{c}} 
  }
  \startdata
  12.67 & 4.7 kpc  & 28$^\circ$ & 62$^\circ$\\
\enddata
  \tablenotetext{a} {Total magnitude (AB) of the exponential disk.}
  \tablenotetext{b} {Disk scale length.}
  \tablenotetext{c} {Inclination angle of the disk.}

\label{tab:expdisk}
\end{deluxetable}

The best fit parameters of the exponential disk are listed in Table \ref{tab:expdisk}.
This exponential disk component dominates in the $i^\prime$ band brightness;
its brightness is more than half of the SDSS $i$ band magnitude, 11.99$\pm$0.05 \citep{Brown2014}.  
The other S$\acute{\rm e}$rsic components are 2--3 magnitude fainter than the disk, thus do not contribute to the
continuum light significantly.
\citet{Bland1991} performed a Fabry-Perot spectroscopy of the central region of NGC~6240 and
found a rotating disk component whose PA and inclination are $\sim$45$^\circ$ and $\sim$70$^\circ$,
respectively.
These values are roughly consistent with the parameters of the exponential disk we fitted.
Thus we conclude that there still is a large stellar disk, which is probably the remnant of 
the primary galaxy of the merger,  in the center of the system.  
We refer to this disk as the ``main stellar disk'' of NGC~6240.

\subsection{Morphology of H$\alpha$ nebula}

\begin{figure*}
\plotone{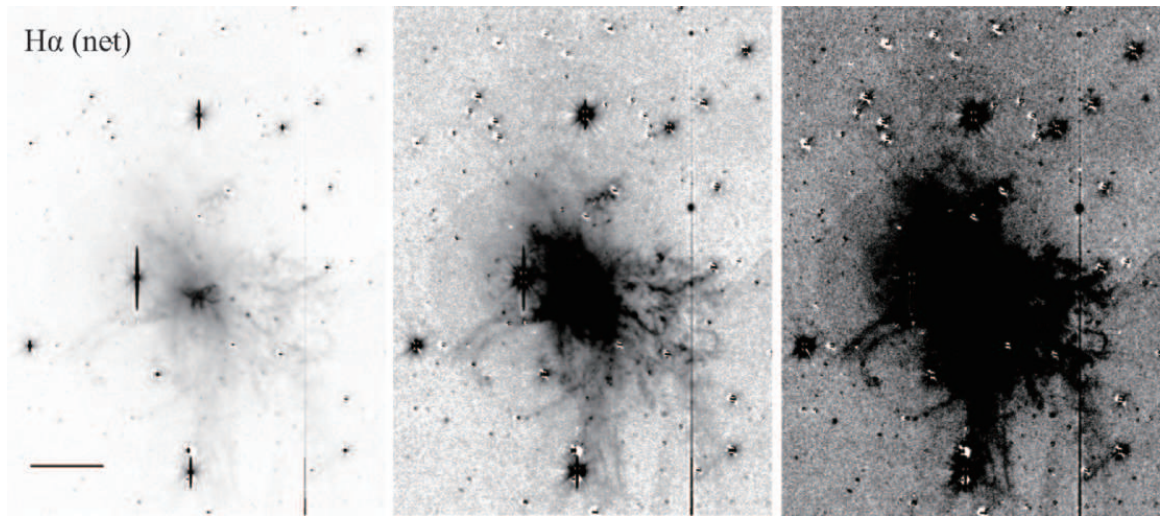}
\caption{Three different contrast versions of the continuum-subtracted 
H$\alpha$+[\ion{N}{2}] (net H$\alpha$+[\ion{N}{2}]) image of NGC~6240.
The scale bar at the bottom left corner of the left panel corresponds to 20 kpc.
Before subtraction, the blooming patterns of bright stars in the $R_{\rm C}$ band image were
removed by linear interpolation of the background.
Some artificial features introduced by this procedure are visible around the bright stars
in this image. 
Note that a circular pattern seen around the bright star 
east-northeast of the center and a large
faint arc pattern seen at the southwest corner of the image 
(these are clearly seen in the middle and right panels) 
are the ghost halos of bright stars.
}
\label{fig:H-alpha}
\end{figure*}

\begin{figure*}
\plotone{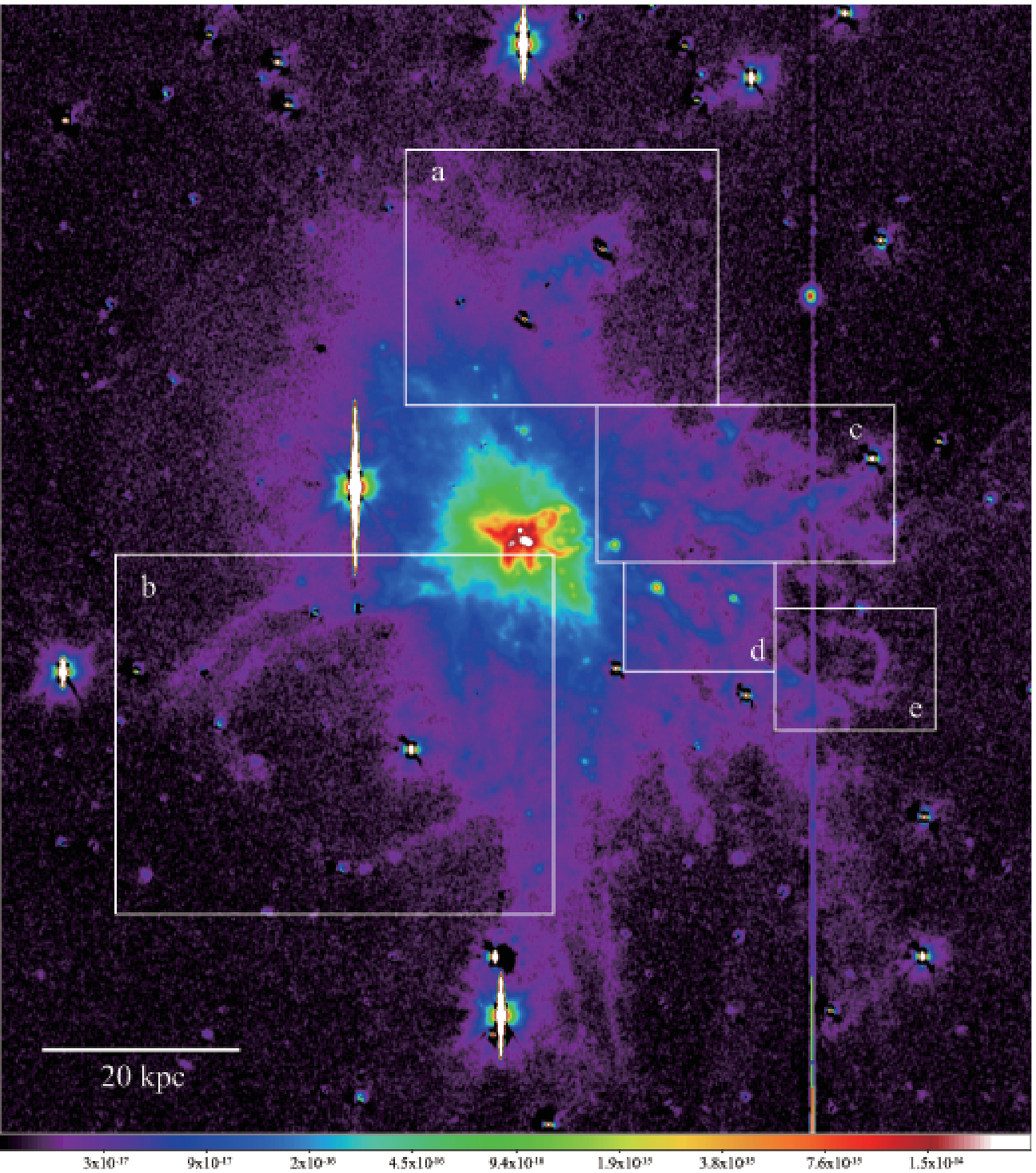}
\caption{False color representation of the net H$\alpha$+[\ion{N}{2}] image.
The unit of the color scale bar is erg~s$^{-1}$~cm$^{-2}$~arcsec$^{-2}$.
Close-up views of the five rectangular regions (a - e) presented here are
shown in Figure \ref{fig:detail}. 
}
\label{fig:H-alpha-color}
\end{figure*}

\begin{figure*}
\plotone{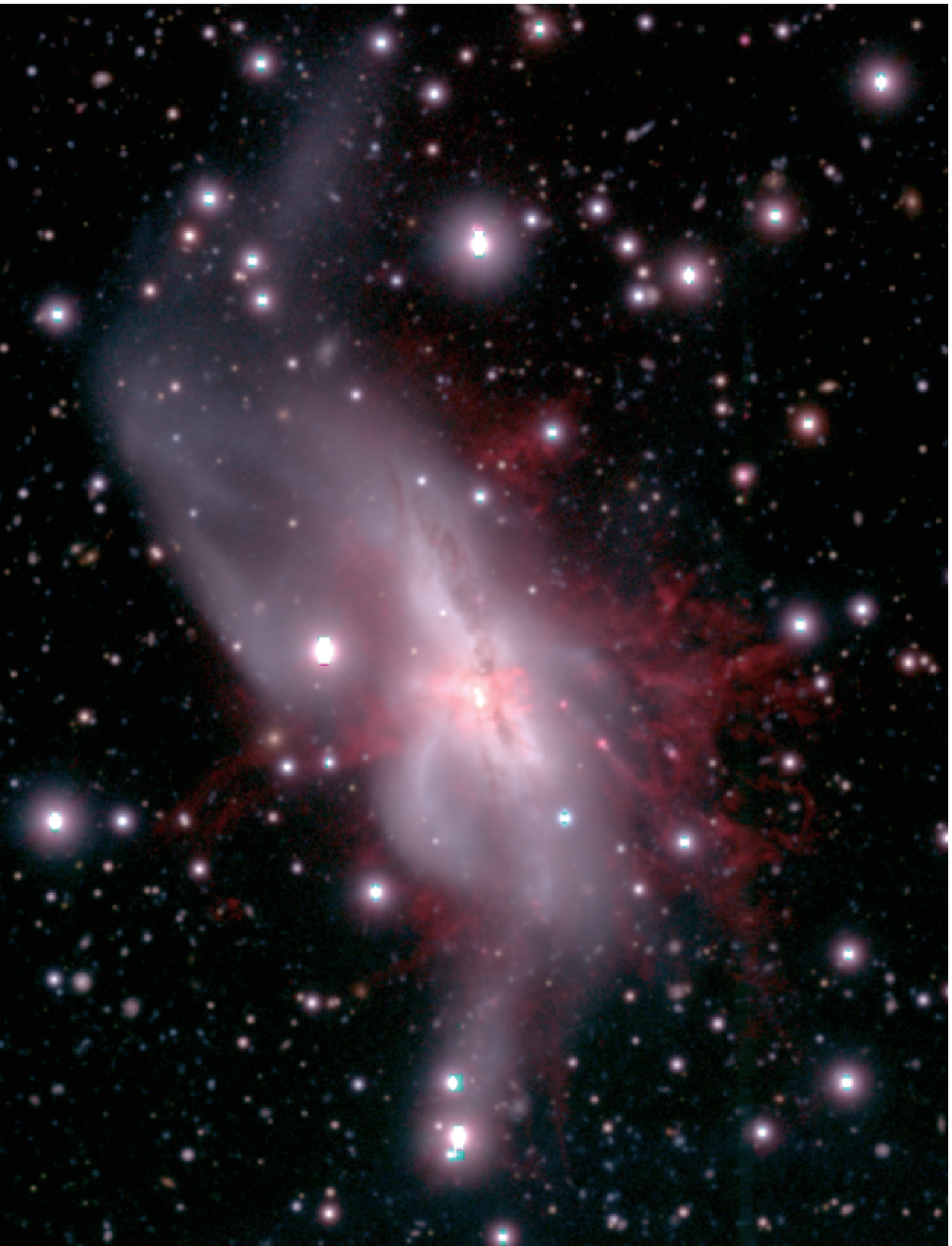}
\caption{
False color image of NGC~6240. The scale for this figure is 105 kpc $\times$ 140 kpc.
North is up and east is left. 
We assigned the $B$ image, $R_{\rm C}$ image, and H$\alpha$+[\ion{N}{2}] image to 
blue, green, and red colors, respectively. 
Highly extended H$\alpha$ emitting gas is clearly seen in the red color.}
\label{fig:colorimg}
\end{figure*}

The net H$\alpha$+[\ion{N}{2}] image obtained in this study revealed the rich and complex structure of
the optical emission line nebula of NGC~6240 (Figures \ref{fig:H-alpha}, \ref{fig:H-alpha-color}, and
\ref{fig:colorimg}).
We refer to this nebula as the ``H$\alpha$ nebula''.
The H$\alpha$ nebula has a diameter of $\sim$90 kpc.
The central ``butterfly'' nebula 
\citep[e.g.][]{Gerssen2004} is surrounded by a relatively bright hourglass shaped region
(``hourglass region''; light blue color in Figure \ref{fig:H-alpha-color}; \citealt{Heckman1987}).
In addition to these well-known features, \citet{Veilleux2003} also
detected a faint optical emission line
region extending out to
over 70 kpc $\times$ 80 kpc around NGC~6240.
Our deep Suprime-Cam observations are the first to
reveal the detailed and complex structure of this extended
component of the H$\alpha$ nebula.

\begin{figure*}
\plotone{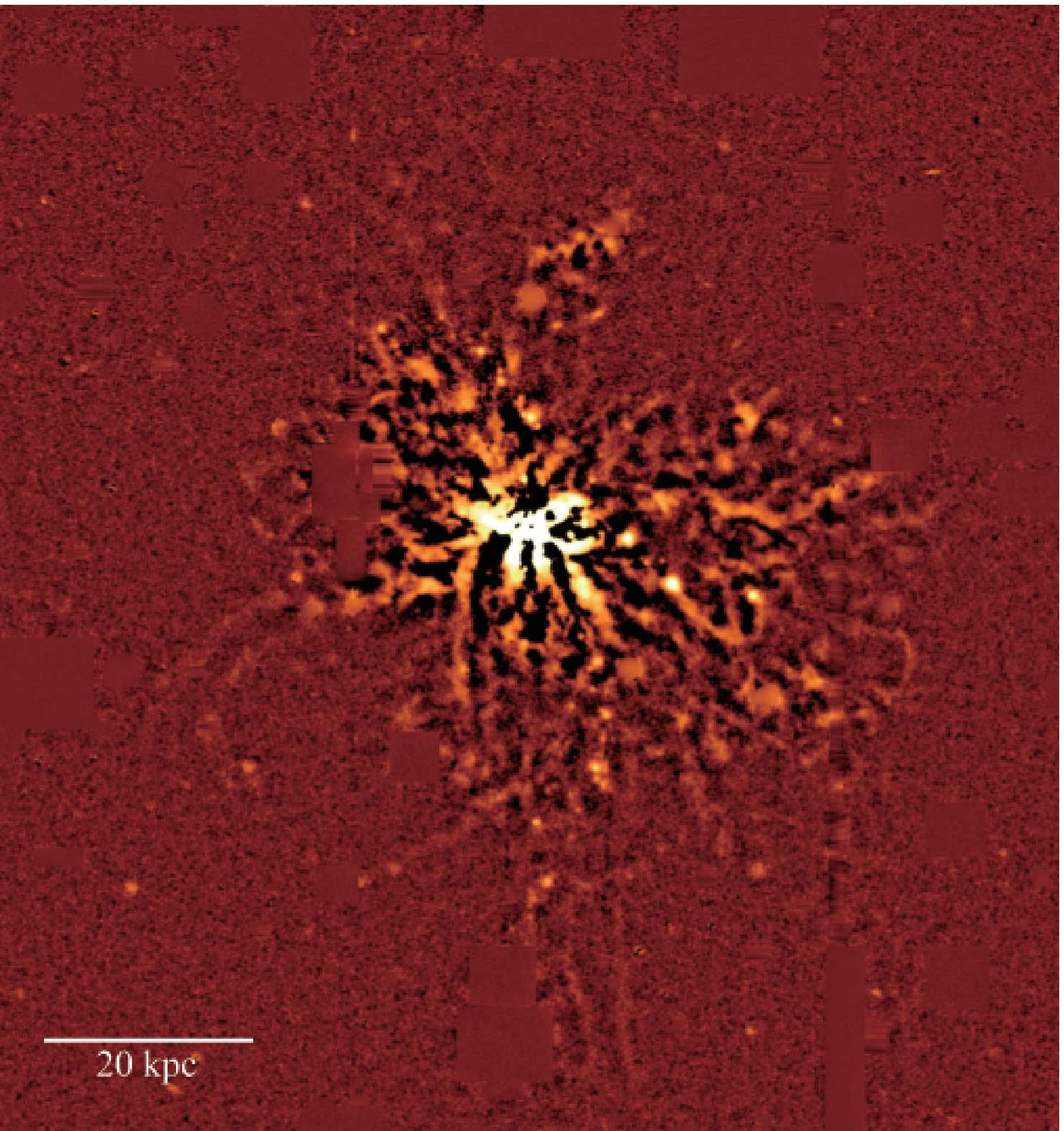}
\caption{Un-sharp masked net H$\alpha$+[\ion{N}{2}] image. 
Bright stars and artifacts are masked before un-sharp masking.
The detailed structure of the filaments can be observed. 
}
\label{fig:unsharp}
\end{figure*}

\begin{figure}
\plotone{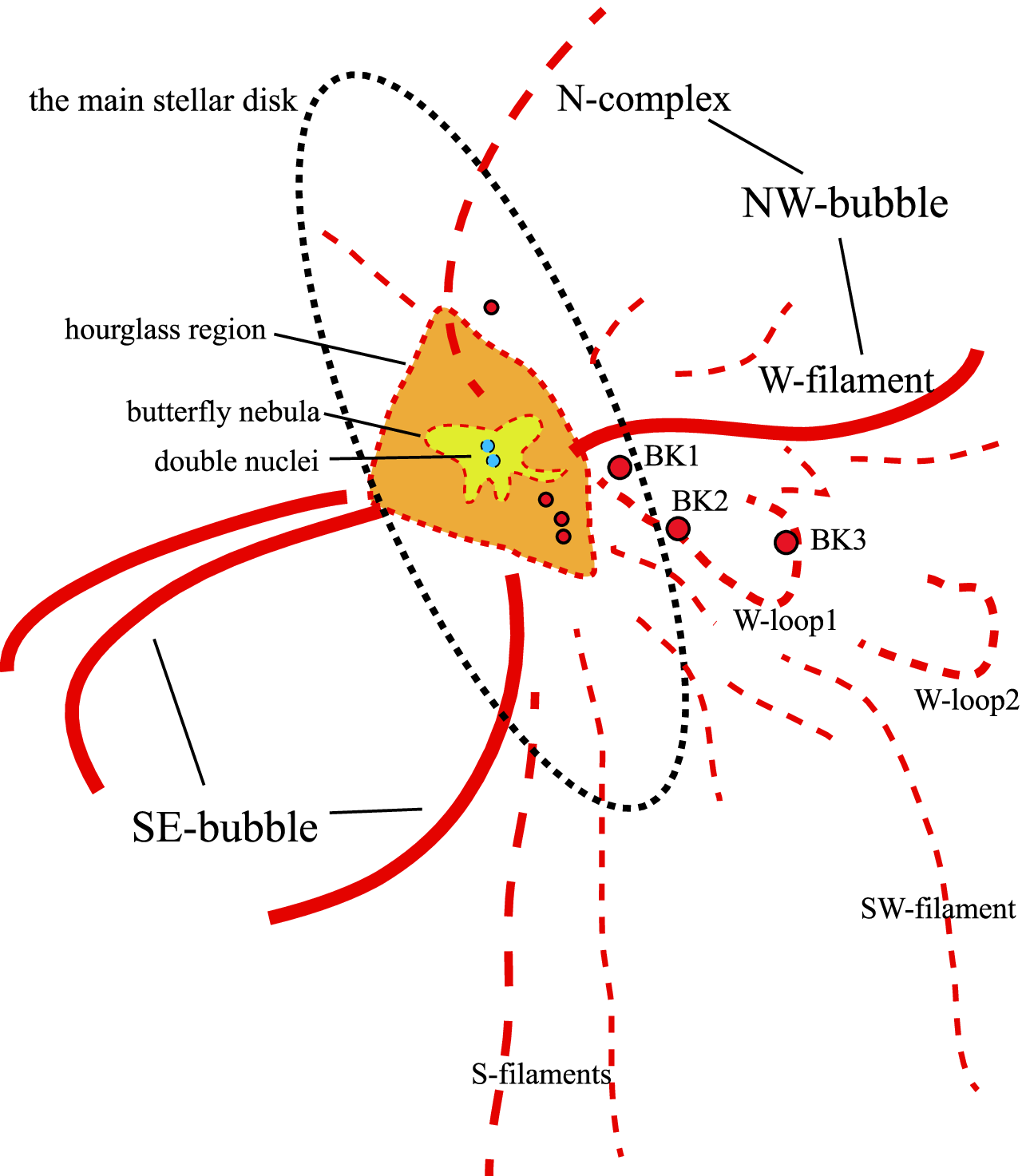}
\caption{Sketch of the H$\alpha$ nebula of NGC~6240.
}
\label{fig:sketch}
\end{figure}

We applied an unsharp masking technique to the net H$\alpha$+[\ion{N}{2}] image
to enhance the filamentary structure
of the H$\alpha$ nebula.
We first masked bright stars and blooming patterns in the H$\alpha$+[\ion{N}{2}]
image.
Second, we smoothed the masked image with a 35 $\times$ 35 pixels 
($\approx$3.4 kpc $\times$ 3.4 kpc) median filter,
then subtracted the smoothed image from the masked image.
The unsharp masked image is shown in Figure \ref{fig:unsharp}.
It is clear that a number of filaments are running radially from the galactic center.
A sketch of the structure of the H$\alpha$ nebula is shown in Figure \ref{fig:sketch}.

We identified several characteristic features in this nebula.
In the following subsections, we describe these features in detail.

\subsubsection{West filament}

The most remarkable feature of the extended H$\alpha$ nebula is a bright, wiggled
filament (``W-filament'') in the western region.
It is prominent in the middle panel of Figure \ref{fig:H-alpha}
(see panel (c) of Figure \ref{fig:detail} for the detailed structure). 
Although the W-filament was previously visible in the narrow-band H$\alpha$ images
taken by \citet{Heckman1987} and \citet{Veilleux2003}, its detailed structure and surrounding 
complex features were only revealed in our H$\alpha$+[\ion{N}{2}] image.
The bright part of the W-filament begins at 18~kpc west from the s-nucleus
and extends to the west-southwest.
The filament turns slightly west-northwest at $\sim$25 kpc and reaches $\sim$36 kpc
from the s-nucleus.
There is a bright cloud to the east of the bright region of the W-filament ($\sim$12 kpc west
from the s-nucleus).
We assume, from a morphological point of view, this could be a part of the W-filament
(Figure \ref{fig:sketch}). 

\subsubsection{Northern complex of clouds and filaments}

\begin{figure*}
\plotone{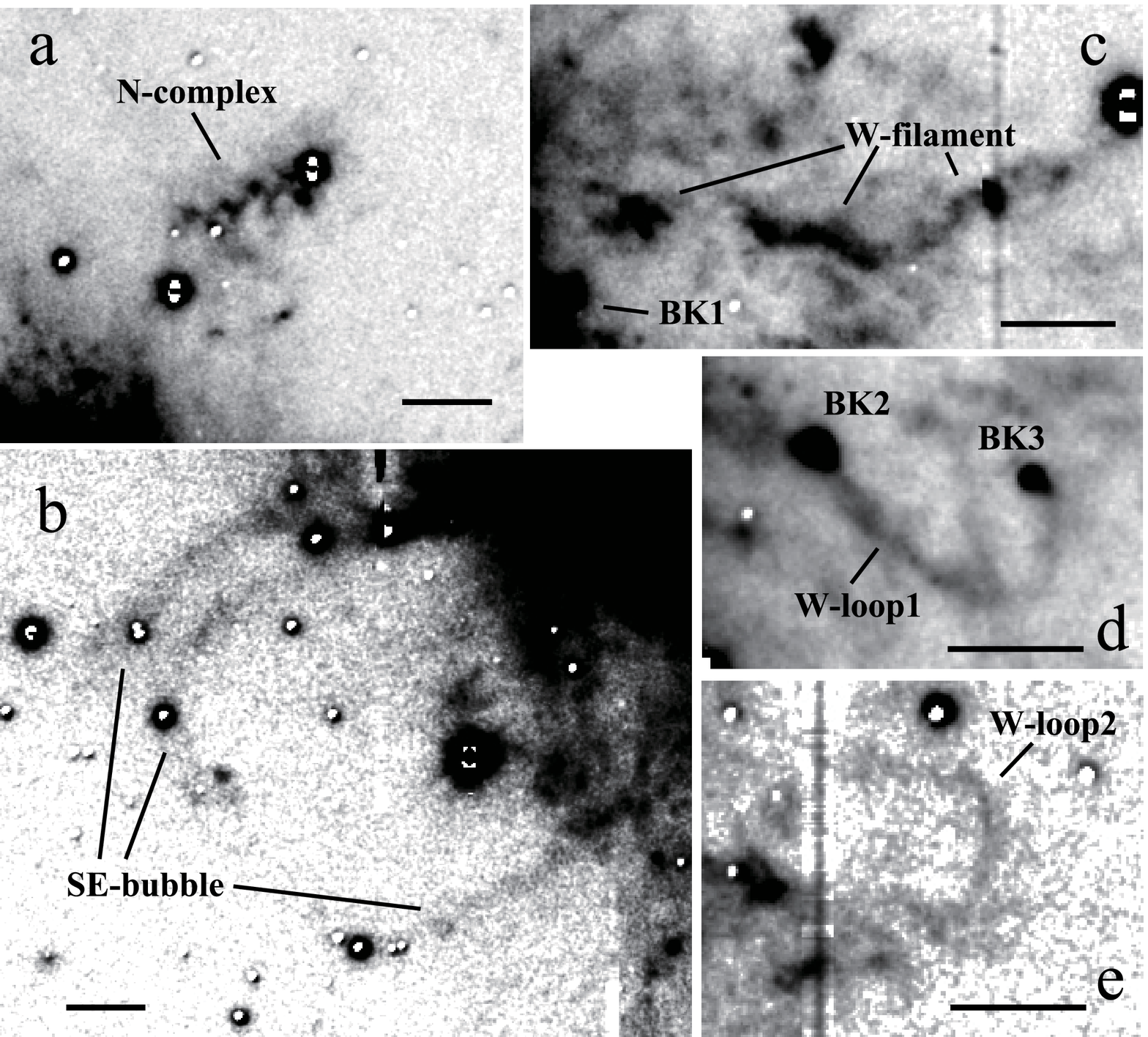}
\caption{Close-up views of the various characteristic features of the H$\alpha$ nebula of NGC~6240.
The scale bars correspond to 5 kpc. 
{\it a.} Northern complex of filaments and blobs (N-complex). 
{\it b.} Southeast giant bubble (SE-bubble). 
{\it c.} Western wiggled bright filament (W-filament).
{\it d.} Bright knots (BK2 and BK3) and loop like structure (W-loop1) connecting these knots.
{\it e.} Southwestern faint loop (W-loop2).
}
\label{fig:detail}
\end{figure*}

On the northern side of the galaxy, there exists a complex of clouds and filaments (``N-complex'': 
Figure \ref{fig:sketch} and panel (a) of Figure \ref{fig:detail}), 
which extends along the PA $\sim$310$^\circ$ from
$\sim$25 kpc north to $\sim$33 kpc north-northwest of the galaxy.
The unsharp masked image indicates that the N-complex is the tip of a large
curved structure connected to the hourglass region (Figure \ref{fig:unsharp}).

\subsubsection{Southeastern and northwestern bubbles}
\label{sec:bubble}

A clear bubble-like structure is visible in the southeastern side of the galaxy 
(``SE-bubble'':
Figure \ref{fig:sketch} and panel (b) of Figure \ref{fig:detail}).
This faint bubble was unknown before this study.
The size and PA of the SE-bubble are $\sim$28 kpc $\times$ 38 kpc and $\sim$135$^\circ$,
respectively.
On the other side of the galaxy, the N-complex and 
W-filament appear to form a large broken bubble
whose size is comparable to the SE-bubble (``NW-bubble'').
The PA of the NW-bubble axis is approximately $-$40$^\circ$.
The axes of these two bubbles are almost perpendicular to the major axis
of the main body of NGC~6240.
The opening angle of the SE-bubble and the NW-bubble is $\sim$70$^\circ$--90$^\circ$ 
(Figures \ref{fig:unsharp} and \ref{fig:sketch}).

\subsubsection{Bright knots and western loops}

The western side of the H$\alpha$ nebula is much brighter and has 
a more complex structure than the eastern side.
In particular, the W-filament and its surroundings have a rich
structure consisting of many filaments and blobs.

There are several bright knots (``BK1''--``BK3'': Figure \ref{fig:sketch}) 
located in a loop-like filament  
extending from the central nebula towards the southwest
(``W-loop1'': Figure \ref{fig:sketch}).
These knots are marginally resolved in our H$\alpha$+[\ion{N}{2}] image; 
the FWHMs of the knots are $\approx$1\farcs 0--1\farcs 1. 
W-loop1 extends almost straight towards the southwest from the center, then
turns to the north at $\sim$23 kpc drawing a loop-like trajectory 
(panel (d) of Figure \ref{fig:detail}).
One of the bright knots (BK3) is located at the tip of this loop.

Another loop (``W-loop2'') is located at $\sim$35 kpc from the center
(panel (e) of Figure \ref{fig:detail}).
The size and shape of W-loop2 are similar to those of W-loop1, but its
surface brightness is much lower than that of W-loop1.

\subsubsection{Southern and southwestern filaments}

There are two streams extending along the north-south direction on the south side of the galaxy
(``S-filaments'' and ``SW-filaments'': Figure \ref{fig:sketch}; see also Figure \ref{fig:unsharp}).
The S-filaments consist of several filaments extending from the center towards the south.
The SW-filaments consist of a number of filaments distributed from 
the region close to W-loop2 to 
$\sim$59 kpc southwest from the s-nucleus.
The extension of the S-filaments is similar to but slightly offset from that of the southern tidal
feature which is the southern end of the S-shaped morphology of the optical continuum emission of NGC~6240.
There is no clear optical continuum counterpart to the SW-filaments.

\subsection{Physical parameters of H$\alpha$ nebula}

\subsubsection{H$\alpha$ surface brightness profile and luminosity}
\label{sec:lum}

\begin{figure}
\plotone{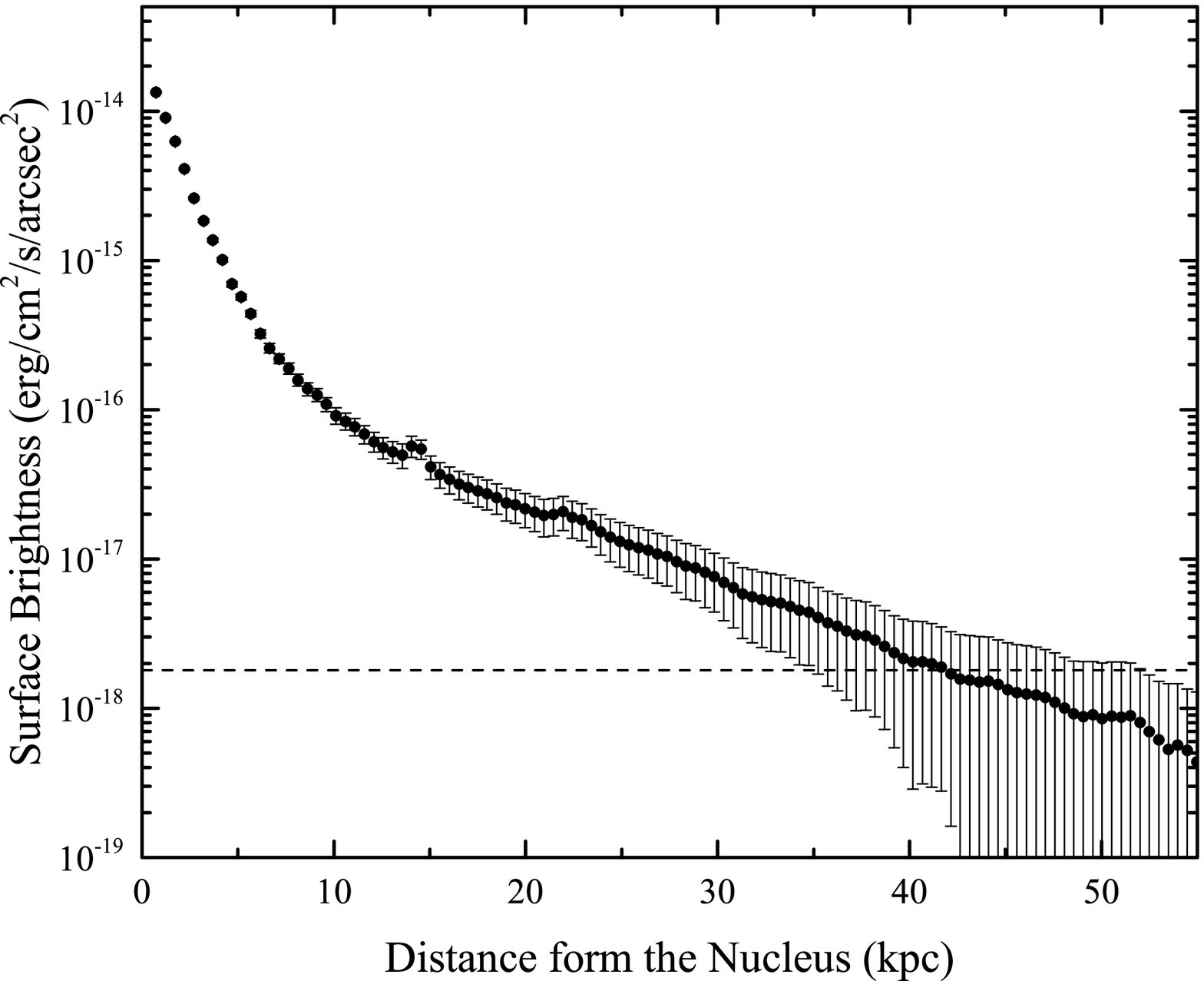}
\caption{Surface brightness profile of the H$\alpha$ emission of the H$\alpha$ nebula.
The contribution of the [\ion{N}{2}]$\lambda\lambda$6548,6583 emission is corrected by
assuming a constant $f_{\rm [N II]6583} / f_{\rm H\alpha}$ ratio of 1 for the entire
region of the nebula (see text).
The dashed line represents the 1$\sigma$ level of the sky background.
}
\label{fig:sb}
\end{figure}

We performed multiple aperture photometry for the H$\alpha$ emission of NGC~6240.
We selected the region further out than 140\arcsec\ ($\sim$70 kpc) from the s-nucleus
for sky background estimation.
Before performing photometry, bright stars and artifacts were masked and 
interpolated by linear functions
around the patterns.
We adopted circular apertures 
whose center was the brightest point of the H$\alpha$+[\ion{N}{2}] image
corresponding to the s-nucleus.
We excluded the very center ($r <1$\arcsec) region in this aperture photometry procedure, 
because the $R_{\rm C}$ band frame that was subtracted from the 
N-A-L671 frame to create the net H$\alpha$+[\ion{N}{2}] image
was saturated at the central 1\arcsec\ region of the galaxy.
The contamination of the [\ion{N}{2}]$\lambda\lambda$6548,6583 emission was corrected
assuming $f_{\rm[N II]6583} / f_{\rm H\alpha} = 1$ \citep{Heckman1990,Keel1990,Veilleux2003}
for the entire region of the nebula.
In addition, we assumed that the foreground extinction is the same as the $R$ band Galactic 
extinction toward NGC~6240;  $A_{\rm R}$ = 0.165 magnitude
\citep{Schlafly2011}.

The surface brightness profile of the H$\alpha$ emission is given in Figure \ref{fig:sb}.
It is clear that the H$\alpha$ surface brightness distribution indicates the core-halo structure
of the nebula, which is similar to the case of the soft X-ray nebula \citep{Huo2004,Nardini2013},
whereas the core is more compact in H$\alpha$.

The total H$\alpha$ flux of NGC~6240 was derived by integrating the fluxes in the regions at
which the H$\alpha$ surface brightness is greater than 2$\sigma$ of the sky background.
We assumed that the H$\alpha$ surface brightness within the central 1\arcsec\ region 
which we excluded in the aperture photometry is the same as that at 1\arcsec\ away from
the s-nucleus.
As mentioned above, we assumed that $f_{\rm[N II]6583} / f_{\rm H\alpha}$ $= 1$
for the whole region of the nebula.
This ratio tends to increase toward the 
outer part of the H$\alpha$ nebula, reaching $\gtrsim$1.4 around the W-filament
\citep{Veilleux2003}.
Thus our method may underestimate the contribution of the [\ion{N}{2}] emission in some parts of
the nebula.
However, since most of the bright regions of the nebula have almost constant 
$f_{\rm[N II]6583} / f_{\rm H\alpha}$ $\sim 1$ \citep{Keel1990,Veilleux2003}, 
the variation of this ratio
does not significantly affect the estimation of the total H$\alpha$ flux.
The resultant total H$\alpha$ flux $f_{\rm H\alpha}$ is $\approx$1.3 $\times$ 10$^{-12}$ erg~s$^{-1}$~cm$^{-2}$.
The H$\alpha$ luminosity $L_{\rm H\alpha}$ is $\approx$1.6 $\times$ 10$^{42}$ erg~s$^{-1}$.

\subsubsection{Electron density, filling factor and mass}
\label{sec:ed}

\begin{deluxetable*}{cccccc}
  \tablewidth{0pt}
  \tablecaption{Electron densities and volume filling factors}
  \tablehead{
    \colhead{$r$ range \tablenotemark{a}} & 
    \colhead{$n_{\rm e, X}$ (rms) \tablenotemark{b}} &
    \colhead{$T_{\rm X}$ \tablenotemark{b}} &
    \colhead{$n_{\rm e, H\alpha}$ (rms) \tablenotemark{c}} &
    \colhead{$f_{\rm v, H\alpha}$} &
    \colhead{$f_{\rm v, H\alpha}$} \\
    \colhead{[kpc]} & 
    \colhead{[cm$^{-3}$]} &
    \colhead{[K]} &
    \colhead{[cm$^{-3}$]} &
    \colhead{case 1) \tablenotemark{d}}&
    \colhead{case 2) \tablenotemark{e}}
  }
  \startdata
     0 $-$ 1.7 & $4.5\times10^{-2}$ & $1.3\times10^7$ & $2.4\times10^{-1}$ &  $1.6\times10^{-5}$ & $2.9\times10^{-5}$ \\ 
   1.7 $-$ 3.4 & $2.3\times10^{-2}$ & $1.1\times10^7$ & $1.2\times10^{-1}$ &  $2.2\times10^{-5}$ & $2.7\times10^{-5}$ \\ 
   3.4 $-$ 7.4 & $6.7\times10^{-3}$ & $8.7\times10^6$ & $5.2\times10^{-2}$ &  $7.9\times10^{-5}$ & $5.8\times10^{-5}$ \\ 
   7.4 $-$ 11 & $3.5\times10^{-3}$ & $8.4\times10^6$ & $2.8\times10^{-2}$ &  $8.9\times10^{-5}$ & $6.5\times10^{-5}$ \\ 
   11 $-$ 15 & $2.4\times10^{-3}$ & $7.8\times10^6$ & $2.0\times10^{-2}$ &  $1.1\times10^{-5}$ & $7.1\times10^{-5}$ \\ 
   15 $-$ 20 & $2.0\times10^{-3}$ & $7.7\times10^6$ & $1.4\times10^{-2}$ &  $8.5\times10^{-5}$ & $4.9\times10^{-5}$ \\ 
   20 $-$ 27 & $1.6\times10^{-3}$ & $8.1\times10^6$ & $1.1\times10^{-2}$ &  $7.5\times10^{-5}$ & $4.7\times10^{-5}$ \\ 
   27 $-$ 39 & $1.4\times10^{-3}$ & $7.2\times10^6$ & $0.7\times10^{-2}$ &  $5.2\times10^{-5}$ & $2.5\times10^{-5}$\\
\enddata
  \tablenotetext{a} {Averaging radial range.}
  \tablenotetext{b} {Nardini et al. (2013).}
  \tablenotetext{c} {RMS electron density of H$\alpha$ emitting gas calculated by emission measure of the H$\alpha$ nebula, assuming a spherical symmetric geometry of the nebula.}
  \tablenotetext{d} {The volume filling factor of H$\alpha$ gas calculated assuming a simple thermal pressure equilibrium with $f_{\rm v, X} = 10^{-2}$ and $T_{\rm H\alpha} = 10^5$ K.}
\tablenotetext{d} {$f_{\rm v, H\alpha}$ calculated assuming a ram pressure equilibrium with $f_{\rm v, X} = 10^{-2}$ and $\eta = V_{\rm s, X} / V_{\rm s, H\alpha} = 10$.}

\label{tab:eden}
\end{deluxetable*}

\begin{figure}
\plotone{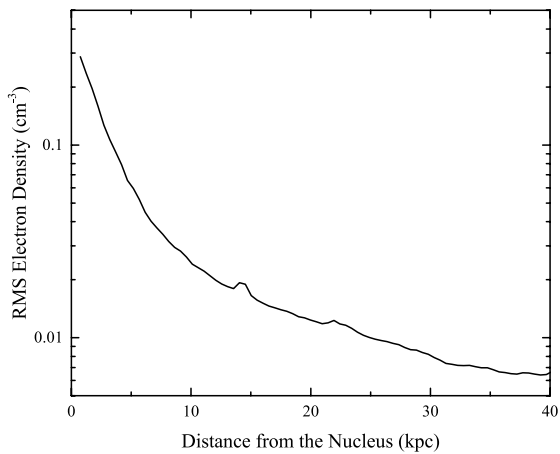}
\caption{Root-mean-square (rms) electron density of the H$\alpha$ nebula 
derived by the emission measure,
assuming a spherically-symmetric distribution of the ionized gas.
}
\label{fig:ne}
\end{figure}

We derived the root mean square (rms) electron density 
$n_{\rm e, H\alpha}$(rms) of the H$\alpha$ nebula assuming Case B
recombination.
We did not, however,  apply an extinction correction to this calculation.  
The surface brightness of H$\alpha$ $SB_{\rm H\alpha}$ is defined as
\begin{equation}
SB_{\rm H\alpha} = \alpha_{\rm B, H\alpha} \cdot h \nu_{\rm H\alpha} \cdot {\rm EM} ,
\end{equation}
where $\alpha_{\rm B, H\alpha}$ and EM are the photon production coefficient of
H$\alpha$ in Case B recombination \citep{Osterbrock2006} and the emission measure,
respectively.
Assuming a fully-ionized pure hydrogen gas 
($n_{\rm e} \approx n_{\rm p}$; $n_{\rm p}$ is proton density) 
and a constant density distribution along the line of sight, 
the EM can be written as ${\rm EM} = n_{\rm e, H\alpha}^2  l$, 
where $l$ is the line of sight length of the nebula.
In the case of spherically symmetric nebula, $l = 2 \sqrt{R^2 - r^2}$, where $R$ and $r$ are
the radius of the nebula and the projected distance from the nebula center, respectively.

Thus, the $n_{\rm e, H\alpha}$(rms) is given by
\begin{equation}
n_{\rm e, H\alpha} ({\rm rms}) = \sqrt{\frac{SB_{\rm H\alpha}}{\alpha_{\rm B, H\alpha} \cdot h \nu_{\rm H\alpha} \cdot l }} .
\end{equation}
The radial profile of the $n_{\rm e, H\alpha}$(rms) distribution is shown in Figure \ref{fig:ne}.

The mass of the H$\alpha$ emitting gas $M_{\rm H\alpha}$ is given by 
\begin{equation}
M_{\rm H\alpha} = \int f_{\rm v, H\alpha}^{1/2} \cdot n_{\rm e, H\alpha} ({\rm rms})\;  m_{\rm H} \; dV ,
\end{equation}
where $f_{\rm v, H\alpha}$ and $m_{\rm H}$ are the volume filling factor of the H$\alpha$ emitting gas
and the mass of a hydrogen atom, respectively.

We estimated the volume filling factor $f_{\rm v, H\alpha}$ by the following procedure.
First, considering the close spatial correlation between the H$\alpha$ nebula and 
the soft X-ray-emitting hot gas (see section \ref{compX} below),
we assumed that the H$\alpha$ gas and X-ray gas are locally coexisting,
Then we considered the following two cases for the coexistent state; 1) a simple thermal pressure equilibrium case and 2)
a ram pressure balance case.

In the case 1), we calculated the local electron densities $n_{\rm e, H\alpha}({\rm local})$ of the H$\alpha$ 
emitting cloud assuming  that the equation
$n_{\rm e, X}({\rm rms}) \cdot f_{\rm v, X}^{1/2} \cdot T_{\rm X} = n_{\rm e, H\alpha}({\rm local}) \cdot T_{\rm H\alpha}$
holds over each radial zone in which \citet{Nardini2013} provided the averaged electron densities 
$n_{\rm e, X}$ and temperatures $T_{\rm X}$ of the hot gas individually.
Here, $f_{\rm v, X}$ is the volume filling factor of the hot gas.
Because the optical emission line filaments of NGC~6240 show clear shock excitation nature in
its emission line ratios \citep{Heckman1990,Keel1990},
we assumed that $T_{\rm H\alpha} = 10^5$ K (typical temperature of optical shock excitation gas
\citep{Shull1979}) for all of the zones in this case.

The case 2) is the one that shock excitation is responsible for the H$\alpha$ emission gas
and X-ray emitting hot gas.
We calculated the $n_{\rm e, H\alpha}({\rm local})$ assuming the ram pressure balance equation \citep{Spitzer1978},
$n_{\rm e, X}({\rm rms}) \cdot f_{\rm v, X}^{1/2} \cdot V_{\rm s, X}^2 = n_{\rm e, H\alpha}({\rm local}) \cdot V_{\rm s, H\alpha}^2 $ ,
where $V_{\rm s, X}$ and $V_{\rm s, H\alpha}$ are the shock velocities at the shock front 
between the hot gas and warm gas, respectively.
The shock velocity required to produce X-ray emitting gas with a temperature of $\sim$10$^7$ K is
$\sim$1000 km~s$^{-1}$ for an adiabatic shock \citep{Hollenbach1979}.
\citet{Heckman1990} found that the averaged [\ion{O}{1}]$\lambda 6300$/H$\alpha$ and 
[\ion{S}{2}]$\lambda\lambda$6717+6731/H$\alpha$ emission line intensity ratios of the optical filaments of 
NGC~6240 are $\sim$0.3 and $\sim$0.8, respectively.
These values are consistent with shock models with the shock velocity $\sim$100 km~s$^{-1}$ \citep[e.g.,][]{Shull1979}. 
Hence, we assume that $V_{\rm s, X}$ and $V_{\rm s, H\alpha}$ are $\sim$1000 km~s$^{-1}$ and $\sim$100 km~s$^{-1}$,
respectively.

We then derived $f_{\rm v, H\alpha}$ by 
$f_{\rm v, H\alpha} = ( n_{\rm e, H\alpha} ({\rm rms}) / n_{\rm e, H\alpha}({\rm local}) )^2$ for the both cases.
In the case 1),  $f_{\rm v, H\alpha}$ is given by
\begin{multline}
f_{\rm v, H\alpha} \approx 10^{-4} \cdot \biggl( \frac{f_{\rm v, X}}{10^{-2}} \biggr)
\biggl( \frac{T_{\rm H\alpha}}{10^5\; {\rm K}} \biggr)^2
\biggl( \frac{T_{\rm X}}{10^7\; {\rm K}} \biggr)^{-2}\\
\biggl( \frac{n_{\rm e, H\alpha}({\rm rms})}{10^{-1}\; {\rm cm^{-3}}} \biggr)^2
\biggl( \frac{n_{\rm e, X}({\rm rms})}{10^{-2}\; {\rm cm^{-3}}} \biggr)^{-2}.
\label{eq:fv1}
\end{multline}
In the case 2), we obtain
\begin{multline}
f_{\rm v, H\alpha} \approx 10^{-4} \cdot \biggl( \frac{f_{\rm v, X}}{10^{-2}} \biggr)
\biggl( \frac{\eta}{10} \biggr)^{-4}
\biggl( \frac{n_{\rm e, H\alpha}({\rm rms})}{10^{-1}\; {\rm cm^{-3}}} \biggr)^2\\
\biggl( \frac{n_{\rm e, X}({\rm rms})}{10^{-2}\; {\rm cm^{-3}}} \biggr)^{-2},
\label{eq:fv2}
\end{multline}
where $\eta = V_{\rm s, X} / V_{\rm s, H\alpha}$. 

The volume filing factor $f_{\rm v, X}$ is not well known, but
the soft X-ray would be strongly enhanced at the upstream of the shocks in the wind
\citep{Strickland2000,Cooper2008,Cooper2009} and the $f_{\rm v, X}$ of such
shocked gas clouds is considerably smaller than 1.
\citet{Strickland2000} found that the typical $f_{\rm v, X}$ of the X-ray gas in a superwind is of order 10$^{-2}$.
Assuming that $f_{\rm v, X} = 10^{-2}$, $T_{\rm H\alpha} = 10^5$ K,
and $\eta = 10$, we found that the resultant volume filling factors are 
orders of 10$^{-4}$--10$^{-5}$ for the both cases (Table \ref{tab:eden}).
Although these values are much smaller than the typical $f_{\rm v}$ of \ion{H}{2} regions or
planetary nebulae in our Galaxy \citep[$\sim$10$^{-2}$--10$^{-3}$; e.g.][]{Osterbrock2006},
they are comparable to the $f_{\rm v}$ derived for outflow/wind gas by
the observations of nearby starburst galaxies and AGNs
\citep{Yoshida1993,Robinson1994,Cecil2001,Sharp2010}.

Under the assumption of spherically symmetric nebula, we obtained
$M_{\rm H\alpha} \sim 5 \times 10^8$ $M_\odot$
using the $f_{\rm v, H\alpha}$ derived above.
The spherical symmetry is, however, too simple assumption for the geometry of the emission line nebula.
Hence, we next considered a more realistic geometry for the H$\alpha$ nebula; 
it has a filamentary structure and
the emission line filaments occupy a fraction $\zeta$ ($\zeta < 1$) along the line of sight of the spherical nebula.
In this case, the rms electron densities are increased by a factor of $\zeta^{-1/2}$.
Since we think that the X-ray gas and H$\alpha$ gas spatially coexist, 
the line-of-sight occupation factor $\zeta$ is the same for the both gas.
Thus the factor $\zeta$ vanishes in the calculation of the volume filling factor (see equations (\ref{eq:fv1}) and (\ref{eq:fv2})).
On the other hand, the total volume of the nebula is decreased by a factor of $\zeta$.
As a result, the $M_{\rm H\alpha}$ alters by a factor of $\zeta^{1/2}$.
Then we obtained $M_{\rm H\alpha} \sim 5 \times 10^8 \cdot \zeta^{1/2}$ $M_\odot$ for this partly occupied filamentary
nebula case.

\subsection{Comparison with X-ray and UV emission}
\label{compX}

\begin{figure*}
\plotone{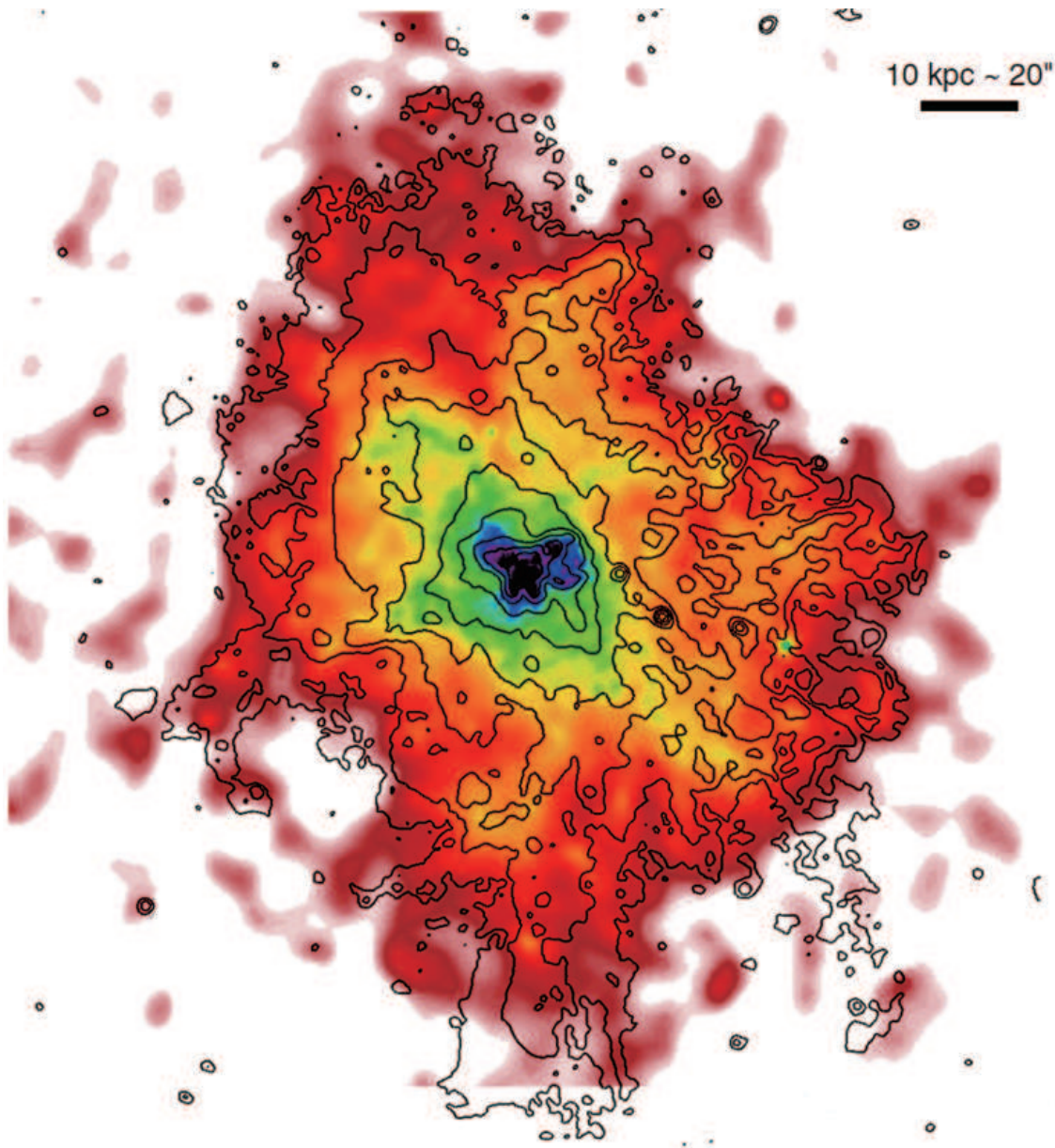}
\caption{H$\alpha$+[\ion{N}{2}] image (contour plot) is overplotted on the adaptively smoothed
soft X-ray image \citep[Figure 4 of][]{Nardini2013}. 
The faintest contour level is $6.2\times10^{-18}$ erg s$^{-1}$ cm$^{-2}$ arcsec$^{-2}$
in H$\alpha$+[\ion{N}{2}],
which corresponds to 2$\sigma$ above the average sky background level.
The interval of the contours is of magnitude 1.0. 
The spatial coincidence between the H$\alpha$+[\ion{N}{2}] and X-ray images is remarkable.
}
\label{fig:X-alpha}
\end{figure*}

We found that the H$\alpha$ nebula shows a very good spatial correlation with 
the highly extended soft 
X-ray-emitting hot gas nebula \citep{Huo2004,Nardini2013}, both in large and small scales.
A tight spatial correlation between the H$\alpha$ and soft X-ray is well known for the butterfly
nebula of NGC~6240 \citep{Komossa2003,Gerssen2004}. 
The overall coincidence between the optical emission line gas and 
the extended soft X-ray gas of NGC~6240
has been previously pointed out by some authors \citep{Veilleux2003,Huo2004}.
Our deep observations, however, allow us to make a more detailed comparison 
between the hot and warm phase gases. 

\begin{figure*}
\plotone{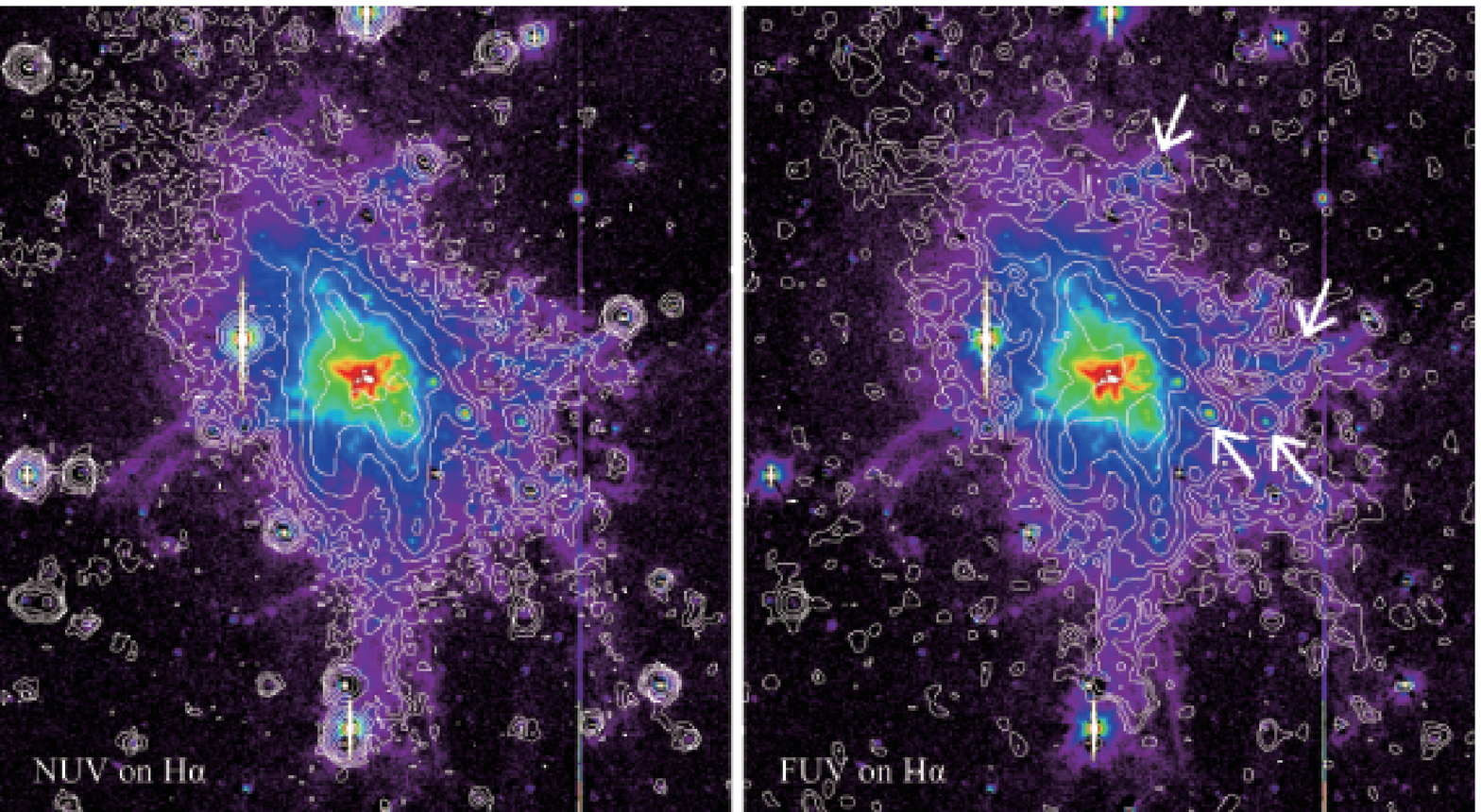}
\caption{Ultraviolet (NUV and FUV) images taken with {\it GALEX} (contour plots) 
are overplotted on 
the false color net H$\alpha$+[\ion{N}{2}] of NGC~6240.
It is remarkable that the western filaments and southern extension of the H$\alpha$ nebula
are spatially coincident with both the NUV and FUV images. 
In particular, the W-filament (panel {\it c} of Figure \ref{fig:detail}) and its surrounding
structures are well traced by the FUV image (shown by arrows in the right panel).
}
\label{fig:uv-H-alpha}
\end{figure*}

We overlaid the contours of the H$\alpha$+[\ion{N}{2}] image on the $\sim$150 ks deep 
soft X-ray image\footnote{This X-ray image is an ``adaptively smoothed''
image \citep[Figure 4 of][]{Nardini2013}; the smoothing scale changes 
with the signal-to-noise
ratio of the original image. See also \url{http://chandra.harvard.edu/photo/2013/ngc6240/ngc6240\_xray.jpg},
and compare the jpg image with our Figure \ref{fig:unsharp} 
to inspect the spatial correlation between the H$\alpha$ nebula and the X-ray nebula
in, more detail.} 
taken with {\it Chandra} \citep{Nardini2013} in Figure \ref{fig:X-alpha}.
The overall shape (a ``diamond shape'') of the faint emission level of
the soft X-ray image is well traced by the H$\alpha$+[\ion{N}{2}] emission.
On smaller scales, there are a number of common features between the
X-ray and H$\alpha$+[\ion{N}{2}], including the central butterfly nebula,
the hourglass region, the N-complex,
the SE-bubble, the W-filament, and W-loop2.
There is also an indication of the counterpart of S-filaments in the X-ray image, 
but the southern tips of the filaments are located outside the X-ray-emitting gas.
There seems to be no clear counterpart to the SW-filaments in the X-ray.
A very tight spatial correlation between the warm 
ionized gas and the hot gas
is maintained from the central region to the edge of the nebula of NGC~6240.

We additionally compared the H$\alpha$ nebula image with 
ultraviolet (NUV and FUV) images of NGC~6240
taken with {\it GALEX}\footnote{{\it GALEX} GR6 data release: \url{http://galex.stsci.edu/GR6/}}
(Figure \ref{fig:uv-H-alpha}).
There also is a good spatial correlation between the H$\alpha$ and the UV images,
particularly in the
western side of the galaxy.
The western UV emission of NGC~6240 is mostly overlapped with 
the western complex of filaments of the H$\alpha$ nebula.
The W-filament and N-complex can be traced in the UV images (Figure \ref{fig:uv-H-alpha}).
The bright knots (BK1 -- BK3) are also visible in the UV images,
and the S-filaments are well traced by the UV emission.
There also are faint UV counterparts for the SW-filaments.

\section{Discussion}

\subsection{Nature of the extended H$\alpha$ nebula}

\subsubsection{Starburst-driven superwind}

We found that the overall morphology of the H$\alpha$ nebula is dominated by
the radially-extended filamentary structure (Figure \ref{fig:unsharp}).
We also identified a large-scale bi-polar bubble (SE- and NW-bubble) along with prominent
loop structures (W-loop1 and 2).
These morphological characteristics indicate that most of the H$\alpha$ nebula
consists of radially-outflowing gas -- the starburst superwind.

The close spatial coincidence between the H$\alpha$ nebula and the extended
soft X-ray gas supports this idea.
Many common features, including the SE-bubble, the N-complex, 
the W-filament, and W-loop2,
are observed in the extended ionized gas of NGC~6240 in both H$\alpha$ and X-ray. 
A similar morphological correlation between the optical emission line region and
the soft X-ray gas has been widely observed in starburst galaxies
\citep{Strickland2000b,Strickland2004,Smith2001,Cecil2002,Veilleux2003,Tullmann2006}.
Such good spatial coincidence has been interpreted as an indication 
that the hot and warm phase gases are heated and excited via
the same mechanism -- probably from shocks in the outflowing gas -- 
over the entire region of 
the nebula. 
Numerical simulations of superwinds have supported this scenario
\citep{Strickland2000,Cooper2008,Cooper2009}.

\begin{figure}
\plotone{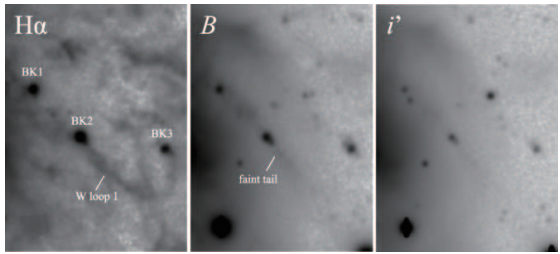}
\caption{Close-up views of the bright knots and W-loop1. 
The net H$\alpha$+[\ion{N}{2}], $B$, and $i^\prime$ images are shown from left to right.
The bright knots are also clearly seen in the broad-band images.
There are faint tails from BK1 and BK2 to the southwestward direction, apparently tracing the
W-loop1 of the H$\alpha$+[\ion{N}{2}] emission.}
\label{fig:knots}
\end{figure}

\subsubsection{Star-forming bright knots}

\citet{Heckman1990} obtained the optical spectra of two of the bright knots 
(BK1 and BK2), and
found that they are active star-forming regions.
Although there are no spectral data for BK3, we suggest that it is also
a star-forming region because the continuum counterparts of 
the bright knots that show
a blue color ($B - i^\prime$ (AB) $\sim -0.3$) are clearly seen 
in our broad-band images (Figure \ref{fig:knots})\footnote{
Although the $B$ and $i^\prime$ images in Figure \ref{fig:knots} were not corrected for emission lines,
we think that the most flux in these bands comes from the stellar continuum. 
In the $B$ band, there are several emission
lines such as H$\beta$, H$\gamma$ and [\ion{O}{3}]$\lambda$5007. We estimated that
the contribution of these
lines to the $B$ band flux is less than 10\% assuming the ordinary Balmer decrement and 
$f_{\rm [OIII]}$/$f_{\rm H\beta}\approx 1$ \citep{Keel1990}.
In the $i^\prime$ band, there is no strong emission lines.}.

It appears that the bright knots are interacting with the outflowing filament, and
the gas in the filament is intercepted by these knots, creating W-loop1.
Tail-like structures are also observed in the broad-band images, in particular,
the tails extending to the southwest are clearly visible
in the $B$ band image (Figure \ref{fig:knots}).
Indications of the tails are also marginally seen in the $i^\prime$ band image.
This feature is recognizable in a Hubble Space Telescope 
(HST) public image of NGC~6240\footnote{
\url{https://www.spacetelescope.org/images/potw1520a/}}.
The head-tail morphology of the knots suggests that the outflow gas 
and condensed gas clouds
in the main stellar disk interact with each other and that its ram pressure 
and/or shock may induce active star formation.
The bright knots are also visible in the UV images supporting the view
that these knots
are the sites of active star formation.
In all probability, the young stellar population observed
in the tails of the bright knots was also formed by
this process.

The S-filaments are well traced by the UV emission.
There also are faint UV counterparts for the SW-filaments.
Because these filaments have either a weak or no spatial correlation 
with the X-ray emission,
we consider that the S- and SW-filaments are not closely related to 
the starburst winds.
These features are more likely to be star-forming tidal tails.

\subsubsection{Extended UV emission}

The UV emission associated with the optical filaments may partly be 
the precursor emission of a shock
within the nebula.
The extended UV emission associated with a superwind has been 
observed in nearby starburst 
galaxies including M82 and NGC~253 \citep[e.g.][]{Hoopes2005}.
In these galaxies, most of the UV emission is thought to be 
from the stellar continuum 
scattered by dust entrained in the wind \citep{Hoopes2005,Hutton2014},
because it is too bright to be provided by shock-heated or photoionized gas in the wind.
It is well-known that NGC~6240 is a dusty system, thus it is possible that the extended 
UV emission of NGC~6240 is mostly derived from light scattered by dust.

To estimate the contribution of dust scattering, we compared the UV intensity
and H$\alpha$ intensity at the W-filament.
We measured the UV and H$\alpha$ intensities 
in a 19\farcs 5 $\times$ 10\farcs 5 rectangular
region at the W-filament (the center of the photometry region is located
54\arcsec\ west and 5\arcsec\ north from the s-nucleus), and determined that
$I_{\rm NUV} =2.0 \times 10^{-14}$ erg~s$^{-1}$~cm$^{-2}$,   
$I_{\rm FUV} =9.3 \times 10^{-15}$ erg~s$^{-1}$~cm$^{-2}$,  
and $I_{\rm H\alpha} = 2.7 \times 10^{-15}$ erg~s$^{-1}$~cm$^{-2}$
(assuming that $f_{\rm [NII]6583}/f_{\rm H\alpha} = 1.4$, \citealt{Veilleux2003}).
The average H$\alpha$ surface brightness in this region was
$1.7 \times 10^{-17}$ erg~s$^{-1}$~cm$^{-2}$~arcsec$^{-2}$.
Both the $I_{\rm NUV}/I_{\rm H\alpha}$ and $I_{\rm FUV}/I_{\rm H\alpha}$ ratios 
-- 7.4 and 3.4, respectively, --
are slightly smaller than, but almost consistent with, values in the outflow 
filaments in M82 and NGC~253 at the same
H$\alpha$ surface brightness level \citep{Hoopes2005}.
The ratios are, however, still larger than the values predicted by the shock model.
This indicates that the UV emission of NGC~6240 is too bright to be 
attributed to shock-heated or
photoionized gas emission in the wind.
Thus, some part of the UV emission is possibly light scattering by dust.  

Alternatively, this emission may reflect extended 
star-forming activity within the filaments.
In this case, gas compression by the interaction of the outflowing gas 
and the interstellar 
medium in the main stellar disk may have induced star formation.
\citet{Nardini2013} found that the X-ray hot gas of NGC~6240 has
a spatially-uniform sub-solar metal 
abundance whose abundance pattern is consistent with Type-II supernova enrichment.
They discussed the age of the nebula using the thermal velocity dispersion 
estimated from the temperature of the hot gas, and concluded that the whole
NGC~6240 nebula could not be 
formed by the recent activity of the nuclear starburst or AGNs.
They suggested that extended continuous star formation from the early phase of 
the merging event would be responsible for the formation of the outer nebula.
The extended UV emission and its spatial coincidence with the H$\alpha$ emission
might support this scenario.

Most of the H$\alpha$ emission of the H$\alpha$ nebula, 
however, probably originates from
shocks in the outflow, because the extended ionized gas shows 
a large [\ion{N}{2}]/H$\alpha$
intensity ratio ([\ion{N}{2}]/H$\alpha$ $\gtrsim$1; \citealt{Keel1990,Veilleux2003}).
Detailed polarimetry and spectroscopy of the extended nebula 
are necessary to qualitatively understand the contribution
of dust-scattering and local star formation to the H$\alpha$ nebula.

\subsection{Ages and durations of superwinds}
\label{sec:age}

\begin{deluxetable*}{lccccccc}
  \tablewidth{0pt}
  \tablecaption{Superwind parameters}
  \tablehead{
    \colhead{Region} & 
    \colhead{Radius} &
    \colhead{$M_{\rm wind}$ \tablenotemark{a}} &
    \colhead{$v_{\rm flow}$} &
    \colhead{SFR} &
    \colhead{Age \tablenotemark{b}} &
    \colhead{Duration \tablenotemark{c}}
\\
    \colhead{ } & 
    \colhead{[kpc]} &
    \colhead{[$M_\odot$]} &
    \colhead{[km s$^{-1}$]} &
    \colhead{[$M_\odot$ yr$^{-1}$]} &
    \colhead{[Myr]} &
    \colhead{[Myr]}
  }
  \startdata
   Butterfly  & 4.5 & $1.4\times10^{8}\;\zeta^{1/2} $ & 700 &  25 & $8.4$ & $1.2\;\xi^{-1} \zeta^{1/2}$ \\ 
   Hourglass & 13 & $5.0\times10^{8}\;\zeta^{1/2} $ &  700 &  25 & $24$ & $4.2\;\xi^{-1} \zeta^{1/2}$ \\ 
   Outer      & 45 & $1.0\times10^{9}\;\zeta^{1/2} $ &  700 &  $\sim 10$ & $84$ & $\sim 20\;\xi^{-1} \zeta^{1/2}$\\
\enddata
  \tablenotetext{a} {Total (H$\alpha$ + X-ray) gas mass entrained and loaded by 
the superwinds assuming that the spherical symmetric geometry of the nebula
and the volume filling factor of the X-ray gas is 0.01.
$\zeta$ is the line-of-sight occupation factor of the nebula (see text).}
  \tablenotetext{b} {Timescale for the gas to reach the end of the region with 
a constant velocity $v_{\rm flow} = 700$ km~s$^{-1}$.}
  \tablenotetext{c} {Duration of the starburst necessary to transfer the gas by 
superwind, where $\xi$ is the thermalization efficiency of
the wind (see text).}

\label{tab:windparam}
\end{deluxetable*}

We discuss the ages and durations of the starburst superwinds of NGC~6240.
From a morphological point of view, we divided the H$\alpha$ nebula into three regions;
the central butterfly nebula \citep{Heckman1987,Keel1990,Lira2002,Gerssen2004}, 
the hourglass region \citep{Heckman1987}, and the outer filaments.
The characteristic radii of the three regions are 
$\sim$4.5 kpc, $\sim$13 kpc, and $\sim$45 kpc, respectively.
We assume that these three regions were formed in the past by 
different starburst/superwind activities, and estimate the ages of 
the regions and the durations of these activities.

First, we estimate the outflow velocity of the winds and derive the dynamical ages of the regions.
\citet{Heckman1990} studied the kinematics of the optical ionized gas of the central
region of NGC~6240 and found that the FWHM of the H$\alpha$ line exceeds 1000
km~s$^{-1}$ within 4--5 kpc from the center \citep[see also][]{Keel1990,Gerssen2004}.
They also found that some parts of the extended region have 
radial velocities of $\sim$500 km~s$^{-1}$.  
We thus adopted a value of 500 km~s$^{-1}$ as the line-of-sight component of 
the outflow velocity of the gas.
Assuming that the typical inclination of the outflow is 45$^\circ$, 
we estimate the outflow velocity of
the H$\alpha$ nebula to be $v_{\rm flow} \sim 700$ km~s$^{-1}$.
We also assume that the outflow velocity is constant over the whole nebula.
The ages of the three regions are thus $\sim$8 Myr, $\sim$24 Myr, and $\sim$84 Myr,
for the butterfly nebula, the hourglass region, and the outer filaments, respectively.

Next, we estimate the durations of the superwind activities that 
formed the three regions based on
the standard superwind model.
The terminal velocity of a mass loading galactic wind is given by
\begin{equation}
V_{\inf} \approx 3000 \times (\xi / \Lambda)^{1/2} \;\; {\rm km\; s^{-1}},
\label{eq:vinf}
\end{equation}
where $\xi$ and $\Lambda$ are the thermalization efficiency and 
mass loading factor of the wind,
respectively \citep{Veilleux2005}.
Substituting $V_{\inf} = v_{\rm flow} = 700$ km~s$^{-1}$ into the equation (\ref{eq:vinf}), 
we find $\Lambda \approx 18 \xi$.
Although the exact value of $\xi$ is unknown and is difficult to estimate, 
it would be expected to lie between 0.1 and 1 \citep{Strickland2000}.
On the other hand, the mass-loss rate from the starburst region 
is scaled by the star formation
rate (SFR) and given by
\begin{equation}
\dot{M_\ast} = 0.26\; (SFR / {\rm M_\odot yr^{-1}}) \; \; {\rm M_\odot\; yr^{-1}}
\end{equation}
\citep{Veilleux2005}.
Then the duration of the superwind $t_{\rm wind}$ is calculated by
$t_{\rm wind} \approx M_{\rm wind} / (\Lambda \times \dot{M_\ast})$, 
leading to
\begin{equation}
t_{\rm wind} \approx 2.1 \times \biggl(\frac{M_{\rm wind}}{\rm 10^8\; M_\odot}\biggr)
\biggl(\frac{\xi}{1.0}\biggr)^{-1} \biggl(\frac{SFR}{\rm 10\; M_\odot yr^{-1}}\biggr)^{-1} \;\; {\rm Myr}
\label{eq:mw},
\end{equation}
where $M_{\rm wind}$ is the total mass of the wind including the loaded mass.

The SFRs are estimated as follows.
Since the butterfly nebula was formed most recent star formation
activities have played important roles in its formation.
We found that there is a wide spread in the estimated values of the current SFR.
A well-used equation for converting the far infrared luminosity 
to the SFR \citep{Kennicutt1998},
$SFR$ $M_\odot$~yr$^{-1}$ $\approx 17$ $(L_{\rm FIR}/10^{11} {\rm erg~s^{-1}})$,  
gives us an SFR of $\sim$120 $M_\odot$~yr$^{-1}$.
This value is clearly overestimated,
because a significant fraction of the bolometric luminosity of
NGC~6240 is known to originate from the AGNs 
and extended star formation other than the current 
nuclear starburst \citep{Lutz2003,Engel2010,Mori2014}.
\citet{Engel2010} found a SFR of $\sim$25 $M_\odot$~yr$^{-1}$ 
for the nuclear starburst
based on the $K$ band luminosity.
On the other hand, \citet{Beswick2001} found a SFR of 83 $M_\odot$~yr$^{-1}$ based on 
the 1.4 GHz radio continuum observation.
\citet{Engel2010} noted that the discrepancy between the above two values arises
from the difference in the star formation histories they used; 
\citet{Beswick2001} assumed
a constant star formation history over 20 Myr, whereas \citet{Engel2010} adopted
a merger-induced increasing star formation scenario.
The morphology of the H$\alpha$ nebula suggests 
an intermittent star formation nature for
the star-forming history of NGC~6240.
In particular, the estimated age of the butterfly nebula is much shorter than
the duration of the star formation
assumed by \citet{Beswick2001}. 
We thus adopt a SFR of 25 $M_\odot$~yr$^{-1}$ for the butterfly nebula.

The age of the hourglass region estimated above ($\sim$24 Myr) is 
almost consistent with 
the starburst age ($\sim$15--25 Myr) estimated by \citet{Tecza2000}.
We thus interpret that the hourglass region was created by this starburst, 
and its SFR was 25 $M_\odot$~yr$^{-1}$, 
which is the value estimated by \citet{Tecza2000}
from the central red stellar population.
We have no suitable observational basis for estimating the SFR for the outer filaments.
Hence, for the outer filaments we tentatively assume a typical SFR
from the local starburst galaxies, $\sim$10 $M_\odot$~yr$^{-1}$.

The H$\alpha$ gas masses of the three regions were determined 
by integrating the H$\alpha$ gas mass in
three concentric regions; $r$ $\leq$ 4.5 kpc, 4.5 kpc $<$ $r$ $\leq$ 13 kpc, and
13 kpc $<$ $r$ $\leq$ 45 kpc. 
The close morphological correlation between the H$\alpha$ emission 
and soft X-ray of NGC~6240
indicates that these two-phase gases are spatially associated with each other.
Therefore, the mass of the soft X-ray gas must be taken into account 
to derive the total mass of the superwinds $M_{\rm wind}$.
\citet{Nardini2013} estimated the total mass of the soft X-ray gas as 
$\sim$10$^{10}$ $M_\odot$
with the assumption that the volume filling factor $f_{\rm v, X}$ is almost unity and
a spherically symmetric geometry of the hot gas nebula.
However, this assumption is not adequate for the superwind.
As we described in section \ref{sec:ed}, $f_{\rm v, X}$ would be of order 10$^{-2}$.
In addition, we introduced the line-of-sight occupation factor $\zeta$ to represent
the filamentary structure of the nebula (see section \ref{sec:ed}). 
Using these factors, we calculated the total wind mass $M_{\rm wind}$.

The ages and durations of the superwinds that formed the three regions derived by
the above calculations are listed in Table \ref{tab:windparam}.
The duration of the superwind is approximately 
one order of magnitude smaller than the
age of each region.
This suggests that the starburst/superwind activity has 
an intermittent nature as suggested
by numerical simulations of galaxy mergers \citep{Mihos1996,Hopkins2006,Hopkins2013}.
It should be noted that the duration of the superwind for the hourglass region (4 Myr) is 
consistent with the starburst duration  ($\sim$5 Myr) estimated by \citet{Tecza2000}.

In the above discussion, we ignored the AGN contribution to the wind evolution.
Currently, there are double active nuclei at the center of NGC~6240 and they play
a non negligible role in energetics in the central region of the galaxy \citep{Elston1990,Komossa2003,Feruglio2013}.
The AGNs are, however, deeply embedded and still not active enough to form the extended hot and warm gas nebula
surrounding the galaxy \citep{Sugai1997,Mori2014}.
Thus the most part of the nebula was probably formed by the starburst activity \citep{Heckman1990,Keel1990,Nardini2013}.
Even in the vicinity of the nuclei, the influence of the AGNs to the surrounding excited gas region
has not been clear yet \citep{Sugai1997,Ohyama2000,Engel2010,Wang2014}.
Therefore, we conclude that the starburst superwind is the prime mover in forming the extended nebula
and the contribution of the AGNs would be negligible except for the central region (a part of the butterfly nebula).

\subsection{Starburst history of NGC~6240}

NGC~6240 is thought to be in the late stages of 
a galaxy merger \citep{Hopkins2006,Engel2010}.
Its morphology indicates that two gaseous spiral galaxies first encountered 
each other $\sim$1 Gyr ago
and their nuclei are now in the final stages of coalescence.
The H$\alpha$ nebula was formed by intermittent starburst events which
occurred within the last 100 Myr.
Here, we propose a possible scenario of the starburst history of NGC~6240 
from the point of view of the formation of the H$\alpha$ nebula 
by starburst superwinds.

A starburst which occurred $\sim$80 Myr ago created 
powerful bi-polar galactic bubbles;
the NW- and SE-bubbles.
At that time, the ISM in the merging system might be 
more abundant in 
the northwestern and western sides of the merging galaxies than
in the eastern and southeastern sides.
A combination of the intermittent nature of the starburst activity 
and the lopsided ISM distribution
formed a rich structure of loops, filaments, or blobs in the western side of the system.
The extended active star formation as indicated by the UV images 
was also preferentially induced in the western side 
as the wind propagated outward from the 
central starburst.
The bright knots observed in the western side of the system were probably
formed by collisions between
the high-speed outflowing gas from the center and dense gas blobs.
The ram pressure of the outflow served to highly condense 
the gas blobs and initiated active star formation.
 
Approximately 20 Myr ago, a short duration starburst ($\sim$4--5 Myr) 
occurred at the center of the system.
This starburst and the subsequent superwind formed 
the central stellar population of NGC~6240 and 
the hourglass region of the H$\alpha$ nebula.  

Active star formation ceased temporarily following the short duration burst
described above.
Several Myr ago, the concentration of gas by 
the continuing approach of the two nuclei 
induced an intense starburst that continues to the present;
and also drove AGN activity.
The SFR may be steadily increasing, 
and has currently reached $\sim$25 $M_\odot$~yr$^{-1}$
\citep{Engel2010}.
This recent starburst and AGN activity formed various features including the butterfly nebula
\citep{Heckman1990,Keel1990,Komossa2003,Gerssen2004}, the central extended
hard X-ray gas \citep{Wang2014}, the molecular outflow 
\citep{Tacconi1999,Nakanishi2005,Iono2007,Engel2010,Feruglio2013,Scoville2015}, 
and the intense H$_2$ and [\ion{Fe}{2}] emission
\citep{Tanaka1991,Werf1993,Sugai1997,Ohyama2003}.

\section{Summary}

We conducted unprecedented deep H$\alpha$ narrow-band imaging 
observations of NGC~6240 using the Subaru Suprime-Cam.
The 6,000-sec Suprime-Cam observation was the first 
to reveal the detailed structure of a highly-extended H$\alpha$ emitting 
nebula (the H$\alpha$ nebula) whose size reaches to $\sim$90 kpc in diameter.
This nebula was probably formed by large-scale energetic outflows
--- or superwinds --- 
driven by starburst events induced by a galaxy--galaxy merger.
In addition to the butterfly-shaped inner nebula, 
which has already been investigated 
by various authors, the outer nebula has a complex structure 
consisting of wiggled narrow filaments, loops, bubbles, and blobs.
We found a huge bi-polar bubble (the SE-bubble plus the NW-bubble), 
which extends 
along the minor axis of the main stellar disk of NGC~6240.
The total H$\alpha$ luminosity is $\approx$1.6 $\times$ 10$^{42}$ erg~s$^{-1}$.

We estimated the volume filling factor $f_{\rm v, H\alpha}$ of the H$\alpha$ nebula assuming a simple
thermal pressure equilibrium and a ram pressure balance between the warm ionized gas and the soft X-ray gas.
The $f_{\rm v, H\alpha}$ is of orders 10$^{-4}$--10$^{-5}$ for the both cases.
We calculated the mass of the H$\alpha$ emitting gas to be 
$M_{\rm H\alpha}$ of $\sim$5 $\times 10^8$ $M_\odot$ using the $f_{\rm v, H\alpha}$.

It is remarkable that the H$\alpha$ extent and morphology are quite similar 
to the X-ray (0.7--1.1 keV) emitting hot gas detected with a deep {\it Chandra}
observation \citep{Nardini2013}.
Such a close spatial coincidence between the optical emission line nebula 
and the hot gas is 
generally observed in starburst superwinds.
This strongly suggests that the same physical mechanism, 
shock-heating, plays an important role 
in the excitation of both phases of the gas at the outer regions of the H$\alpha$ nebula. 

The H$\alpha$ nebula also has a close spatial correlation
with the UV emission in the western
region of the galaxy.
Some bright star-forming knots are located in the western region.
These facts suggest, therefore, that star-forming activity was induced by
the interaction of the outflow and the ISM,
and that the interaction was probably enhanced due to a lopsided gas distribution 
formed by the merger.

The morphology and surface brightness distribution of the H$\alpha$ nebula suggest
that there were probably three major episodes of starburst/superwind activities 
that formed the nebula.
The most extended part of the nebula was initiated $\sim$80 Myr ago.
Multiple intermittent starbursts in this era created 
many outflow filaments and clouds extending
$\sim$40 kpc out from the center.
At the same time, the outflowing gas interacted with the ISM 
and induced extended star formation
in the main stellar disk.
An intense short duration starburst then occurred $\sim$20 Myr ago, forming
the hourglass region of the H$\alpha$ nebula.
NGC~6240 entered a new active phase $\sim$1 Myr ago.
The galaxy is now in the final stages of the merger and 
its activity is increasing.

\acknowledgments

We thank the anonymous referee for the careful reading and constructive
suggestions.
We are grateful to the staff of the Subaru telescope for their kind
help with the observations.
This work was in part carried out using the data obtained by a collaborative
study on the cluster Abell~1367.
This study was in part carried out using the facilities at
the Astronomical Data Analysis Center (ADAC), National Astronomical
Observatory of Japan. This research made use of NASA's
Astrophysics Data System Abstract Service, NASA/IPAC Extragalactic
Database, GALEX GR6 database, and SDSS DR9 database.
This work was financially
supported in part by Grant-in-Aid for Scientific Research 
No.23244030, No.15H02069 from the Japan Society for the Promotion of Science
(JSPS), No.24103003 from the Ministry of Education, Culture, Sports, 
Science and Technology (MEXT), and
MOST grant 104-2112-M-001-034-.


\clearpage
\newpage

\appendix

\section{A. Color conversion from SDSS to the Suprime-Cam filter system}

We adopted a color conversion from SDSS to the Suprime-Cam
filter system with HPK CCDs as 
\begin{equation}
SDSS - Suprime = \sum c_i (SDSS color)^i,
\end{equation}
where the coefficients and the color range of the stars are
given in Table \ref{tab:conv}.
The coefficients are estimated using the method given in \citet{Yagi2013},
with the model quantum efficiency curve of the CCDs.

We used stars of magnitude $17<r<21$ in SDSS DR9.
The K-correct \citep{Blanton2007} v4 offset 
for SDSS\footnote{\url{http://howdy.physics.nyu.edu/index.php/Kcorrect}} 
was applied; $m_{AB}-m_{SDSS}=$ 0.012, 0.010 and 0.028 for $g$, $r$, and $i$, 
respectively.
The number of stars is approximately 60 - 200,
and the estimated rms error 
around the best fit ranged from 0.05 to 0.07.

\begin{deluxetable}{cccccccccccccc}
  \tablewidth{0pt}
  \tablecaption{The coefficients of color conversion from SDSS to Suprime-Cam}
  \tablehead{
    \colhead{SDSS-Suprime} & 
    \colhead{SDSS color} &
    \colhead{range} &
    \colhead{$c_0$} &
    \colhead{$c_1$} &
    \colhead{$c_2$} &
    \colhead{$c_3$} &
    \colhead{$c_4$} &
    \colhead{$c_5$} &
    \colhead{$c_6$} &
    \colhead{$c_7$}
  }
  \startdata
$g-B$ &  $g-r$ & $-0.6<g-r<0.6$    & $-0.031$ & $-0.124$ & 0.070 & $-0.415$ & $-0.213$ & 0.881 & ...   & ...  \\
$r-R$ &  $r-i$ & $-0.6<r-i<0.6$    & 0.006 & 0.312 & $-0.064$ & $-0.152$ & 1.601 & $-1.262$ & $-7.990$ & 9.992\\
$r-NA671$ &  $r-i$ & $-0.6<r-i<0.3$ & 0.020 & 0.536 & 0.021 & $-0.381$ & $-0.722$ & ... & ... & ...  \\
$i-i$     &  $r-i$ & $-0.6<r-i<1.8$ & $-0.006$ & 0.089 & 0.019 & $-0.016$ & ... & ... & ... & ...  \\
\enddata

\label{tab:conv}
\end{deluxetable}

\section{B. Correction of the contaminated emission line flux for the net H$\alpha$+[\ion{N}{2}] image}

The use of the scaled $R_{\rm C}$ frame as an off-band frame leads 
an over-subtraction of the continuum, because the H$\alpha$+[\ion{N}{2}]
emission is non-negligibly contaminated in the scaled $R_{\rm C}$ frame.
The optical emission line nebula of NGC~6240 primarily consists of shock-heated gas
\citep[e.g.][]{Heckman1990,Keel1990},
thus the [\ion{O}{1}]$\lambda$6300 and [\ion{S}{2}]$\lambda\lambda$6717,6731 lines
must also contribute towards contaminating the emission line flux in the $R_{\rm C}$ frame.
Assuming that the emission line intensity ratio 
$f_{\rm [O I]+[S II]} / f_{\rm H\alpha+[N II]}$ is
constant over the whole region of the nebula, we can write the H$\alpha$+[\ion{N}{2}] flux in the net H$\alpha$+[\ion{N}{2}] frame $f_{\rm H\alpha+[N II]}^\prime$ as
\begin{equation}
f_{\rm H\alpha+[N II]}^\prime = (1 - \beta) f_{\rm H\alpha+[N II]},
\label{eq:f}
\end{equation}
where $\beta$ is the contamination factor in the $R_{\rm C}$ frame,
$f_{\rm H\alpha+[N II]}^\prime$ is the H$\alpha$+[\ion{N}{2}] flux uncorrected for over-subtraction of 
line emission from the $R_{\rm C}$ frame, and $f_{\rm H\alpha+[N II]}$ is that corrected 
for this effect.
Here, we assumed that the filter transmissions are the same for all the emission lines,
for simplicity.
We then derived the scaling factor $c$ in the following equation by subtracting 
the net H$\alpha$+[\ion{N}{2}] frame multiplied by $c$ from the scaled $R_{\rm C}$ frame 
so that the central butterfly nebula component disappears in the subtracted frame,
\begin{equation}
c\;f_{\rm H\alpha+[N II]}^\prime = \beta\;  f_{\rm H\alpha+[N II]}
\label{eq:c}.
\end{equation}
Equations (\ref{eq:f}) and (\ref{eq:c}) give $f_{\rm H\alpha+[N II]} = (1+c) f_{\rm H\alpha+[N II]}^\prime$.
We found that the best scaling factor $c$ was 0.23.
Thus the correction factor for the H$\alpha$+[\ion{N}{2}] flux of the net H$\alpha$+[\ion{N}{2}] image
is 1.23.

\end{document}